\documentclass[aps,prc,twocolumn,showpacs,superscriptaddress,nofootinbib,nolongbibliography]{revtex4-2}  
\usepackage{graphicx}  
\usepackage{dcolumn}   
\usepackage{bm}        
\usepackage{amssymb}   
\usepackage{amssymb,amsmath,amsfonts} 
\usepackage[breaklinks=true,debug=true]{hyperref}
\usepackage{braket}    
\usepackage{multirow}
\usepackage{adjustbox}
\usepackage{gnuplottex}
\usepackage[T1]{fontenc}
\usepackage[utf8]{inputenc}
\usepackage{color}
\usepackage{dsfont}
\usepackage{tikz}
\usepackage{filecontents}
\usepackage{pgfplots}
\usepackage{slashbox}
\usepackage{array}
\usepackage{rotating}

\usepackage[para,online,flushleft]{threeparttable}

\usepgfplotslibrary{groupplots}
\pgfplotsset{compat=newest}
\usepgfplotslibrary{units}
\usetikzlibrary{backgrounds}

\newcolumntype{M}[1]{>{\centering\arraybackslash}m{#1}}

\begin{document}


\title{Shell-Model Description of the Isospin-Symmetry-Breaking Correction to Gamow-Teller $\beta$-Decay Rates and Their Mirror Asymmetries}

\author{L.~Xayavong} 
\email{xayavong.latsamy@yonsei.ac.kr} 
\affiliation{Department of Physics, Yonsei University, Seoul 03722, South Korea}
\author{Y.~Lim}
\email{ylim@yonsei.ac.kr}
\affiliation{Department of Physics, Yonsei University, Seoul 03722, South Korea}
\vskip 0.25cm  
\date{\today}

\begin{abstract} 

The isospin-symmetry breaking correction, denoted as $\delta_C$, is introduced for the first time within the shell-model framework to the nuclear matrix element of Gamow-Teller transitions. $\delta_C$ is separated into two components: the isospin mixing term, $\delta_{C1}$, induced by the Coulomb and nuclear charge-dependent forces in the effective Hamiltonian, and the radial mismatch term, $\delta_{C2}$, arising from differences between proton and neutron realistic wave functions. Consequently, the refinement strategy developed for superallowed $0^+\rightarrow 0^+$ Fermi transitions is applied to Gamow-Teller transitions as well. It is demonstrated that, to a given precision level, the shell model calculation of $\delta_C$ converges much faster than the calculation of the transition matrix element. 
Furthermore, higher-order correction terms are investigated and considered for consistent study of our works. 
Various interesting properties of the leading correction terms are discovered within the two-level model and parentage expansion of the one-body transition densities in angular momentum and isospin spaces. One such property is the dependence of $\delta_{C1}$ on the isospin admixture amplitude, $\alpha$, starting from the first order, while the same model yields $\delta_{C1} = \alpha^2$ for Fermi transitions. Numerical calculations are performed for a number of $p$- and $sd$-shell nuclei using existing phenomenological effective Hamiltonians and their isospin nonconserving versions. A realistic Woods-Saxon potential is employed to simulate the radial mismatch effect, with its depth and length parameters re-fitted case-by-case to fix separation energies and charge radii, respectively, when available. The calculated $\delta_C$ values are then utilized to evaluate the mirror asymmetry of Gamow-Teller transition strengths, which are compared with available experimental data and other theoretical calculations. 
Due to the refined fitting procedure of the Woods-Saxon potential parameters and the improved convergence as a function of intermediate state number, our results show better agreement on average compared to those of Smirnova and Volpe [Nucl. Phys. {\bf A 714}, 441 (2003)]. 

\end{abstract}

\maketitle

\section{Introduction}

The atomic nucleus has long served as a low-energy laboratory for testing the fundamental symmetries underlying the Standard Model of electroweak theory. The possibility of assigning nuclear states with well-defined quantum numbers offers a unique opportunity to investigate the effects of various selection rules in electroweak interactions. Among nuclear weak processes, the superallowed $0^+ \rightarrow 0^+$ Fermi transition of isotriplets ($T=1$) provides the cleanest probe of the vector coupling constant, $G_V$, and the subsequent $V_{ud}$ element of the Cabibbo–Kobayashi–Maskawa (CKM) quark-mixing matrix. Furthermore, the nuclear data on $V_{ud}$, together with the data on $V_{us}$ obtained from Kaon decays, can be used to verify the unitarity of the CKM matrix. A violation of this condition may be indicative of the existence of a fourth generation of elementary particles.  
These investigations not only contribute to our understanding of electroweak interactions but also offer the potential to unveil new physics beyond the Standard Model. 

The superallowed $0^+\rightarrow0^+$ Fermi transition of  isotriplets can only connect the isobaric analog states of nuclei with isospin projection quantum numbers of $T_z=-1$, $0$, and $1$, which are located around the $N=Z$ line of the nuclear chart. As a pure vector ($V$) process, its corrected $\mathcal{F}t$ value (denoted as $\mathcal{F}t^{0^+\rightarrow 0^+}$) can be defined as follows: 
\begin{equation}\label{Ft}
\begin{array}{ll}
    \mathcal{F}t^{0^+\rightarrow 0^+} &= f_Vt(1+\delta_R')(1-\delta_C^V+\delta_{NS}^V) \\[0.1in]
    &= \displaystyle\frac{K}{\mathcal{M}_F^2G_F^2V_{ud}^2(1+\Delta_R^V)}, 
\end{array}
\end{equation}
the top-right term of Eq.~\eqref{Ft} contains the experimental $f_Vt$ value, which is the product of the statistical rate function and partial half-life, along with 
theoretical corrections $\delta_C^V$, $\delta_R'$, and $\delta_{NS}^V$. These corrections account for isospin-symmetry breaking (ISB), transition-dependent, and nuclear structure-dependent radiative effects, respectively. The bottom-right term of Eq.~\eqref{Ft} consists of only nucleus-independent quantities: $K$, a combination of fundamental constants~\cite{HaTo2020}; $G_F$, the Fermi constant~\cite{MuLan}; $\Delta_R^V$, another radiative correction term that depends solely on the type of weak current governing the transition under consideration; and $\mathcal{M}_F$, the isospin-symmetry Fermi transition matrix element, given by $\mathcal{M}_F^2=T(T+1)-T_{zi}T_{zf}=2$. It is also noteworthy that $G_V=G_F|V_{ud}|$, and based on the conserved-vector-current (CVC) hypothesis, $G_V$ must remain unaltered in the nuclear medium. 
Therefore, the constancy of $\mathcal{F}t^{0^+\rightarrow 0^+}$ for various superallowed $0^+\rightarrow 0^+$ Fermi transitions of isotriplets serves as a test of the CVC hypothesis. Furthermore, the correlation between $\mathcal{F}t^{0^+\rightarrow 0^+}$ and the average inverse decay energy can be attributed to the presence of the scalar term in the weak hadronic current. 

On the other hand, the superallowed $0^+\rightarrow 0^+$ Fermi transition of isotriplets provides a promising tool for studying nuclear structure. For 15 cases, $f_Vt$ has been measured with a precision of 0.1~\% or better, making the overall uncertainty of $\mathcal{F}t^{0^+\rightarrow 0^+}$ currently dominated by the theoretical corrections, particularly $\delta_C^V$ and $\delta_{NS}^V$, which vary significantly from nucleus to nucleus. Recently, the radiative corrections $\delta_{NS}^V$ and $\Delta_R^V$ have been studied using the dispersion relation theory~\cite{particles4040034,PhysRevC.107.035503}. Although this new formalism for $\delta_{NS}^V$ has yet to be implemented within a reliable nuclear structure model, it has been found that the $\gamma W$-box treatment used by Towner~\cite{TOWNER1992478} 
did not consider the dependence on $\beta$-particle energy, and the use of quenching factors for the axial and magnetic couplings appears to be misinterpreted. 

Although $\delta_C^V$ is principally induced by the Coulomb interaction, which is well-known, its magnitude for the superallowed $0^+ \rightarrow 0^+$ Fermi transition of isotriplets is generally about 1\,\% or smaller—seemingly below the current precision levels of any nuclear structure models. Therefore, obtaining a meaningful result for $\delta_C^V$ from the first-principle nuclear structure calculation is challenging. As an empirical method to go beyond the theoretical precision limit, one may impose as many experimental constraints onto the calculation of $\delta_C^V$ as possible. Towner and Hardy~\cite{ToHa2008} employed this refinement technique within a shell model framework, separating the correction as $\delta_C^V \approx \delta_{C1}^V + \delta_{C2}^V$, where $\delta_{C1}^V$ accounts for the isospin mixing within the shell model configuration space induced by the presence of an isospin nonconserving term in the effective Hamiltonian, and $\delta_{C2}^V$ accounts for the mismatch between proton and neutron radial wave functions. To calculate $\delta_{C1}^V$, they applied an effective isospin nonconserving Hamiltonian that must reproduce low-lying $0^+$ levels, as well as the displacement energy between the initial and final states, mainly determined by the $b$ and $c$ coefficients of the isobaric multiplet mass equation. The calculated values of $\delta_{C1}^V$ were then scaled with the energy separation between the isobaric analog state and the strongest admixed state of the daughter nucleus, following the perspective of the two-level mixing model. The final $\delta_{C1}^V$ values were taken as the average of the scaled and unscaled values, while their differences and the spreads due to the use of different isospin nonconserving effective Hamiltonians were treated as statistical uncertainties~\cite{ToHa2008,HaTo2020}. For the calculation of $\delta_{C2}^V$, they employed Woods-Saxon radial wave functions, whose depth or strength of additional surface term and length parameters were readjusted to match the calculated separation energies and charge radii of the mother nucleus to the corresponding experimental data. The $\delta_{C2}^V$ uncertainties were mainly evaluated from the experimental uncertainties on charge radius data and the different fitting procedures of Woods-Saxon potential parameters. Among the existing calculations of $\delta_C^V$, the shell model with Woods-Saxon radial wave functions provides excellent agreement with the Standard Model predictions, including the CVC hypothesis, CKM unitarity, and the $\mathcal{G}$-parity conservation, which prohibits the existence of weak scalar and tensor currents~\cite{ToHa2010,Smirnova2003}. Since $\delta_C^V$ cannot be directly obtained from experiments, the consistent shell model approach should be thoroughly examined by investigating other nuclear weak processes where this theoretical correction is expected to be larger. 

Persistent efforts have also been made to study the superallowed $0^+\rightarrow 0^+$ Fermi transition of isoquintets ($T=2$)\cite{PhysRevC.81.055503, Bhattacharya, glassman2019superallowed}. This process involves very exotic proton-rich nuclei ($T_z=-2$) and typically decays to an excited state above the proton emission threshold of the daughter nuclei. Since $\delta_{C2}^V$ is very sensitive to the weakly bound effect, the total correction $\delta_C^V$ for this process is larger than that for the Fermi transition of a lower isospin multiplet, including the mirror $\beta$ decay of isodoublets ($T=\frac{1}{2}$)\cite{PhysRevC.107.015502}. The shell model calculation of $\delta_C^V$ for $^{32}$Ar (the best case) appears to agree with the experimentally extracted value, despite not including radiative corrections. However, the experimental uncertainties on $f_Vt$ for this process are still too large to perform a precision test of the standard electroweak theory. 

Generally, extracting weak interaction parameters using Gamow-Teller (GT) transitions is more challenging due to quenching caused by the renormalization of the operator $\bm{\sigma}\bm{\tau}_\pm$ in the nuclear medium~\cite{GT}. This effect is fundamentally a consequence of the non-conservation of the weak axial-vector ($A$) current. However, the $f_At$ ratio of a mirror pair of pure GT transitions is expected to be free from quenching effects and can serve as a clean probe for nuclear structure models. Moreover, the remaining asymmetry in the $f_At$ values of mirror GT transitions, after correcting for nuclear structure and transition-dependent radiative effects, could be attributed to the presence of the induced tensor current~\cite{Smirnova2003}. The ISB contribution to the GT matrix element can be larger than that in Fermi transitions, as the operator $\bm{\sigma}\bm{\tau}_\pm$ can change both the spin and isospin of the nuclear states by one unit, while $\bm{\tau}_\pm$ can only increase or decrease the isospin projection quantum number. Therefore, this axial-vector process is a good candidate for validating the shell model approach for $\delta_C^A$. 

Numerous shell-model studies of the ISB effects on GT transition rates have been carried out for a number of $p$- and $sd$-shell nuclei, with special emphasis on the mirror asymmetry in GT strengths~\cite{Smirnova2003, WILKINSON1972289, BARKER1992147, TOWNER1973589, PhysRevLett.27.1018}. However, these calculations were all directly performed for the nuclear matrix element $M_{GT}$ without introducing $\delta_C^A$. The additional complexity caused by the admixture of isospin mixing and radial mismatch contributions makes such direct calculations more computationally intensive than those through $\delta_C^A$. To reach a desired precision, the calculation of $M_{GT}$ typically requires thousands of intermediate states, while the calculation through $\delta_C^A$ usually needs only a few hundred. This point will be discussed in more detail in the following section.

The aim of the present work is mainly to generalize the shell-model formalism of $\delta_C^A$ to GT transitions. The formula developed so far for Fermi transitions~\cite{xayavong2022higherorder} may not be readily applicable for axial-vector processes due to the differences in their transition operators and the corresponding selection rules. Thus, some important structural behaviors of the leading terms of $\delta_C^A$ for GT transitions are discussed in comparison with those obtained for Fermi processes. In particular, the study of $\delta_{C1}^A$ within the two-level mixing model, as well as the isospin dependence and the impact of deeply bound orbits on $\delta_{C2}^A$ are addressed. Additionally, the higher-order terms of $\delta_C^A$ that were investigated for Fermi transitions in Ref.~\cite{xayavong2022higherorder} are now exactly calculated. Recent developments for the calculation of $\delta_{C2}^V$, including the generalization of the charge radius formula and the self-consistent adjustment of the Woods-Saxon depth or strength of surface-peaked term and length parameters\cite{XaNa2018}, are also implemented. 

This article is organized as follows. In Section~\ref{aa}, we describe the contributions of ISB as well as those due to radiative corrections and physics beyond the Standard Model to the mirror asymmetry in GT transition strengths. In Section~\ref{mixing}, we present the generalized shell model formalism for $\delta_{C1}^A$ and employ the two-level mixing model as a tool to derive an analytic relation for the isospin mixing amplitude and some further interesting properties of the correction term specific to GT transitions. We also provide large-scale shell model calculations using existing isospin nonconserving effective interactions. Section~\ref{radial} is devoted to $\delta_{C2}^A$, including its generalized formalism, investigation of the isospin-dependent properties, sensitivity to deeply bound orbits, and the numerical calculations. The calculations of the higher-order ISB correction terms are given in Section~\ref{x}. The impact of our results on mirror asymmetry and the Standard Model implications are discussed in Section~\ref{c}. Finally, our summary and concluding remarks are given in the last section. 

\section{Mirror asymmetry of Gamow-Teller transition strengths}\label{aa} 

Within the Standard Model framework, a pure GT transition is uniquely governed by the axial-vector part of the weak hadronic current. Therefore, one can customarily define the corrected $\mathcal{F}t$ for GT transition (denoted as $\mathcal{F}t^{GT}$) similar to Eq.~\eqref{Ft}, namely:
\begin{equation}\label{ft}
\begin{aligned}
 f_At(1+\delta_{R}')& (1-\delta_{C}^A + \delta_{NS}^A) \\
&= \frac{K}{\mathcal{M}_{GT}^2G_F^2q_A^2g_A^2V_{ud}^2(1+\Delta_R^A)},
\end{aligned}
\end{equation}
where $G_F$ and $g_A$ represent the Fermi and axial-vector coupling constants, respectively. It is worth noting that the axial-vector current is not conserved, implying that $g_A\ne 1$. The most precise value of $g_A$ was extracted from the $\beta$ decay of the free neutron~\cite{PhysRevLett.122.242501}. $q_A$ denotes the quenching factor used to compensate for the systematic deviation of shell model predictions of GT matrix elements from experimentally deduced values~\cite{USDab}. It should be noted that this compensation may not be needed if one performs an ab-initio calculation, such as the no-core shell model or in-medium similarity renormalization group~\cite{GT}. 

The ISB correction for GT transitions can be defined as $M_{GT}^2=\mathcal{M}_{GT}^2(1-\delta_C^A)$, where $\mathcal{M}_{GT}$ denotes the isospin-symmetry GT matrix element. Following Ref.~\cite{xayavong2022higherorder}, $\delta_C^A$ is customarily written as: 
\begin{equation}\label{cx}
\displaystyle\delta_C^A = \sum_{i=1}^6 \delta_{Ci}^A ,
\end{equation}
where $\delta_{C1}^A$ and $\delta_{C2}^A$ are of leading order (LO), similar to those given in Eq.~\eqref{Ft}. The higher-order terms ($\delta_{Ci}^A, i\in[3,6]$) were found to be negligibly small for the superallowed $0^+\rightarrow 0^+$ Fermi $\beta$ decay of isotriplets. The shell model description of $\delta_C^V$ has been well-established and widely studied. Numerous authors have discussed several important properties of this correction for the superallowed $0^+\rightarrow 0^+$ transitions of isotriplets~\cite{ToHa2008, XaNa2022, OrBr1985} and isoquintets~\cite{Bhattacharya}, as well as the neutron-decay analogy of mirror nuclei~\cite{Severijns2008}. 

\begin{center}
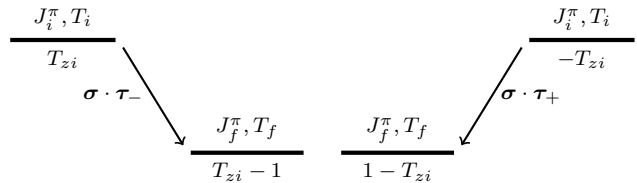
\begin{figure}[ht!]
  \begin{tikzpicture}
  \draw[ultra thick] (-4.4,1.5) -- (-3,1.5) node[above,midway] {\footnotesize$J_i^\pi,T_i$} node[below,midway] {\footnotesize$T_{zi}$};
  \draw[ultra thick] (3.9,1.5) -- (2.5,1.5) node[above,midway] {\footnotesize$J_i^\pi,T_i$} node[below,midway] {\footnotesize$-T_{zi}$};
  \draw[ultra thick] (-2,0) -- (-0.5,0) node[above,midway] {\footnotesize$J_f^\pi,T_f$} node[below,midway] {\footnotesize$T_{zi}-1$};
  \draw[ultra thick] (1.5,0) -- (0,0) node[above,midway] {\footnotesize$J_f^\pi,T_f$} node[below,midway] {\footnotesize$1-T_{zi}$};
  \draw[thick,->] (-2.9,1.4) -- (-2.1,0.1) node[midway,left] {\footnotesize$\bm{\sigma}\cdot\bm{\tau}_-$};
  \draw[thick,->] (2.4,1.4) -- (1.6,0.1) node[right,midway] {\footnotesize$\bm{\sigma}\cdot\bm{\tau}_+$};
  \end{tikzpicture}
\caption{Schematic representation of a mirror pair of Gamow-Teller transitions.}
\label{fx}
\end{figure}
\end{center}

Unlike Fermi transitions, the isospin-symmetry GT matrix element, $\mathcal{M}_{GT}$, generally depends on nuclear structure. Consequently, the corrected $\mathcal{F}t^{GT}$ is no longer a universal constant. Moreover, the factor $q_A$ is not an experimental quantity; instead, it was empirically extracted by assessing the global trend of GT strengths from a systematic study~\cite{GT,USDab}. For these reasons, it would be impossible to extract $g_A$ or $V_{ud}$ via the expression~\eqref{ft} to an acceptable precision level. Nevertheless, the mirror ratio, $(f_At)^+/(f_At)^-$, does not depend on $\mathcal{M}_{GT\pm}$ and $q_A$. Based on Eq.~\eqref{ft} and retaining only LO terms in the total theoretical corrections\footnote{Note that this leading order approximation is used only to facilitate our theoretical interpretation. However, all terms are included in our numerical calculations.}, one can deduce: 
\begin{equation}\label{asym}
\begin{array}{ll}
\delta & \displaystyle=\frac{(f_At)^+}{(f_At)^-}-1 \\[0.14in]
& \displaystyle\approx (\delta_R'^{-}-\delta_R'^{+}) + (\delta_{NS}^{A-}-\delta_{NS}^{A+}) + (\delta_{C}^{A+}-\delta_{C}^{A-}),
\end{array}
\end{equation}
here the $+$ ($-$) label corresponds to the $\beta^+$ ($\beta^-$) emission represented by Fig.~\ref{fx}. 
Notably, this phenomenon is only discernible within regions characterized by 
light and medium-mass regions, typically up to the lower part of the $pf$ shell. In heavier mass regions, the $\beta^+$ emitters of a given mirror pair of GT transitions become too exotic or even unbound against emission of protons. It is evident from the expression~\eqref{asym} that the asymmetry $\delta$ is strongly sensitive to the differences in the correction terms between the mirror GT transitions. More importantly, the theoretical uncertainties on these differences are smaller than those on the individual corrections. 
In general, the radiative corrections are not included in the studies of this weak process, except for the work of Towner\,\cite{TOWNER1992478} considering a few $p$-shell nuclei. Likewise, our present study will focus solely on the contribution arising due to the ISB corrections. By substituting Eq.~\eqref{cx} into Eq.~\eqref{asym}, we obtain: 
\begin{equation}
\delta^{ISB} \approx \sum_{i=1}^6 \delta_i^{ISB} = \sum_{i=1}^6 \left( \delta_{Ci}^{A+} - \delta_{Ci}^{A-} \right), 
\end{equation} 
where $\delta_i^{ISB}$ represents the contribution of $\delta_{Ci}^A$ from the mirror GT transitions. In general, one can expect a positive $\delta^{ISB}$ value due to the stronger Coulomb repulsion in proton-rich nuclei compared to their mirror neutron-rich partners. Among the available experimental data on $\delta$, a negative value was found only for the pair of $^{28}$P($\beta^+$)$^{28m}$Si/$^{28}$Al($\beta^-$)$^{28m}$Si, which disagreed remarkably with the previous shell model calculation~\cite{Smirnova2003}.

As mentioned in the introduction, the beyond Standard Model structure of weak interactions can also contribute to the asymmetry of GT transition strength, particularly the presence of an induced second-class tensor term in the axial-vector current which is not invariant under the $\mathcal{G}$-parity transformation. 
The $\mathcal{G}$-parity violation contribution to the $\delta$ asymmetry is expected to be a linear function of the sum of maximum energy releases in both mirror transitions~\cite{Smirnova2003}. 

\section{Isospin mixing correction}\label{mixing} 
\subsection{General formalism}

The shell model formalism, described in Ref.\cite{xayavong2022higherorder} for Fermi transitions, can be extended for GT transitions by replacing the operator $\bm{\tau}_\pm$ with $\bm{\sigma}\bm{\tau}_\pm$ and by relaxing the angular momentum selection rule accordingly. We notice that we use the bar $\bm{\sigma}\bm{\tau}_\pm$ operator for GT transitions.
Among the six terms of the ISB correction in Eq.\eqref{cx}, $\delta_{C1}^A$ can be expressed in closed form as follows:
\begin{equation}\label{c1}
\delta_{C1}^A = \frac{2}{\mathcal{M}_{GT}} \sum_{k_ak_b} \theta_{ab}^{\lambda}\xi_{ab} D(abfi\lambda),
\end{equation}
where the labels $f$ and $i$ denote the final and initial many-particle states, respectively, and $k_{a/b}$ stands for the set of spherical quantum numbers $(nlj)$ of the respective adjacent single-particle states $a$ and $b$. 

\begin{table}[ht!]
\caption{Analytical expressions for the function $\theta_{ab}^{(\lambda=1)}$. It should be noted that the Gamow-Teller transition selection rule requires no change in the orbital angular momentum, which implies that $l_b=l_a=l$.} \label{tb:theta}
\begin{ruledtabular}
\begin{tabular}{c|c|c}
\backslashbox{$j_a$}{$j_b$} & $ l+\frac{1}{2}$ & $ l-\frac{1}{2}$ \\
\hline
$ l+\frac{1}{2}$ & $\displaystyle\left[ \frac{2(l+1)(2l+3)}{2l+1} \right]^\frac{1}{2}$ & $\displaystyle\left[ \frac{8l(l+1)}{2l+1} \right]^\frac{1}{2}$ \\
\hline
$ l-\frac{1}{2}$ & $\displaystyle-\left[ \frac{8l(l+1)}{2l+1} \right]^\frac{1}{2}$ & $\displaystyle-\left[ \frac{2l(2l-1)}{2l+1} \right]^\frac{1}{2}$ 
\end{tabular}
\end{ruledtabular}
\end{table}

The quantity $D(abfi\lambda)$ represents the deviation of the one-body transition density calculated using an isospin non-conserving effective shell-model Hamiltonian, $\text{OBTD}(abfi\lambda)$, from the corresponding isospin-symmetry value, $\text{OBTD}^T(abfi\lambda)$, namely:
\begin{equation}
D(abfi\lambda) = \text{OBTD}^T(abfi\lambda) - \text{OBTD}(abfi\lambda).
\end{equation}

The term $\text{OBTD}(abfi\lambda)$ is defined as:
\begin{equation}\label{obtd}
\text{OBTD}(abfi\lambda) = \frac{ \braket{f|| [a_b^\dagger \otimes \tilde{a}_a]^{(\lambda)} ||i} }{ \sqrt{2\lambda + 1} },
\end{equation}
where $\lambda=1$ is the rank of the GT transition operator, and the double bar denotes the reduction in angular momentum. On the contrary, a single-particle basis and many-particle wave functions with definite isospin must be employed for the calculation of $\text{OBTD}^T(abfi\lambda)$. 
We note that for the superallowed $0^+\rightarrow 0^+$ Fermi transition of isotriplets, $D(abfi\lambda)$ is always small enough to be considered as a perturbation in the calculation, so that the higher order terms in this quantity can be neglected. In general, this property does not hold for GT transitions as will be seen in the following subsections. 

The isospin component $\xi_{ab}$ in Eq.~\eqref{c1} is given by:
\begin{equation}
\xi_{ab} = \braket{\tau_a|\bm{\tau}\pm|\tau_b} = \left\{
\begin{array}{ll}
1 & \text{for} \hspace{0.1in} \tau_b=\tau_a\mp 1, \\[0.18in]
0 & \text{otherwise,}
\end{array}
\right.
\end{equation}
where $\tau_{a/b}$ denotes the isospin projection quantum number of the single-particle state $a/b$. Our convention for isospin is that $\tau=-\frac{1}{2}$ for protons and $\frac{1}{2}$ for neutrons. 

The function $\theta_{ab}^{\lambda}$ represents the spin-angular component of the single-particle matrix element of the GT transition operator and is defined as: 
\begin{equation}\label{6j}
\begin{array}{ll}
\displaystyle \theta_{ab}^{\lambda} & = \displaystyle (-1)^{l_b+j_b+\frac{1}{2}+\lambda}\sqrt{6(2j_a+1)(2j_b+1)} \\[0.18in] 
& \displaystyle \times\left\{
\begin{array}{lll}
\displaystyle\frac{1}{2} & j_a         & l_b \\
j_b         & \displaystyle\frac{1}{2} & \lambda
\end{array}
\right\} \delta_{l_al_b}. 
\end{array}
\end{equation}

From Table~\ref{tb:theta}, it is evident that the magnitude of $\theta_{ab}^{\lambda}$ increases almost linearly with the orbital angular momentum $l$. For a given value of $l$, $\theta_{ab}^{\lambda}$ reaches its maximum when $j_b=j_a=l+\frac{1}{2}$ (transition between spin-up states), and its minimum when $j_b=j_a=l-\frac{1}{2}$ (transition between spin-down states). Moreover, the sign of $\theta_{ab}^{\lambda}$ is consistently negative when the final state is a spin-down state, specifically $j_a=l-\frac{1}{2}$. Furthermore, it is apparent that the magnitude of $\theta_{ab}^{\lambda}$ remains invariant when reversing the direction of the transition. Consequently, a potential cancellation in $\delta_{C1}^A$ could arise when the sign of $D(abfi\lambda)$ remains constant upon interchanging spin-up and spin-down states. To illustrate, in the context of the $p$-shell model space, Eq.~\eqref{c1} is reduced to: 
\begin{equation}
\begin{array}{ll}
    \delta_{C1} &= \displaystyle \frac{2}{\mathcal{M}_{GT}} \left\{ 4\sqrt{3} \left[ (p_\frac{1}{2}\to p_\frac{3}{2})-(p_\frac{3}{2}\to p_\frac{1}{2}) \right] \right. \\[0.1in]
    &\displaystyle \left. + \frac{2}{\sqrt{6}} \left[ \sqrt{10}(p_\frac{3}{2}\to p_\frac{3}{2}) - (p_\frac{1}{2}\to p_\frac{1}{2}) \right] \right\}. 
\end{array}
\end{equation}
where $(b\to a)$ is shorthand for $D(abfi\lambda)$. 
Hence, the function $\theta_{ab}^{\lambda}$ can exert influence over both the sign and magnitude of $\delta_{C1}^A$. Clearly, this effect is not attributed to nuclear forces, but instead, it is related to the sum rule of GT transition itself. Further examination of $\delta_{C1}^A$ will be provided in the subsequent subsection within the framework of the two-level mixing model. We remark that the function $\theta_{ab}^{\lambda}$ for Fermi transition ($\lambda=0$) is always positive because of the requirement of $j_a=j_b$. 

\subsection{Two-level mixing approximation}

To further understand the properties of $\delta_{C1}^A$, we assume that the initial state of GT transitions possesses a definite isospin, denoted as $\ket{i}=\ket{\omega_iT_iT_{iz}}$, whereas the final state exhibits an isospin admixture, as illustrated by the equation:  
\begin{equation}\label{final}
\ket{f} = \alpha_0\ket{\omega_0T_0T_{fz}} + \sum_{x=1}^\infty \alpha_x\ket{\omega_xT_xT_{fz}},
\end{equation}
where $\omega$ represents all the necessary quantum numbers for labeling nuclear many-body states, apart from angular momentum and isospin. The total angular momentum quantum is conserved by strong interactions, so it is omitted from the above expression. 
Within Eq.~\eqref{final}, the state $\ket{\omega_0T_0T_{fz}}$ corresponds to the final state in the isospin-symmetry limit of the considered GT branch. The normalization condition implies the following:
\begin{equation}
\alpha_0 = \sqrt{1-\sum_{x=1}^\infty\alpha_x^2}.
\end{equation}

According to the first-order perturbation theory, the isospin-mixing amplitude $\alpha_x$ can be written as:
\begin{equation}\label{amp}
\alpha_x = \frac{ \braket{\omega_xT_xT_{fz}|V_{INC}|\omega_0 T_0T_{fz}} }{E_x-E_0},
\end{equation}
where $V_{INC}$ represents the isospin nonconserving component of the final nucleus Hamiltonian. $E_x$ and $E_0$ are the eigenvalues of the respective basis states. It is evident from Eq.~\eqref{amp} that although the isospin mixing is due to the perturbation $V_{INC}$, its degree of influence strongly depends on the energy separation between the admixed states, $E_x-E_0$, which is primarily determined by the isospin-conserving component of the Hamiltonian.

Using the final state given in Eq.~\eqref{final} and neglecting the radial mismatch effect, the corresponding GT transition matrix element is derived as follows:
\begin{equation}\label{eq1}
M_{GT} = \alpha_0\mathcal{M}_0 + \sum_{x=1}^\infty \alpha_x \mathcal{M}_x,
\end{equation}
where $\mathcal{M}_0=\braket{\omega_0T_0T_{fz}|\bm{\sigma}\bm{\tau}_\pm|\omega_iT_iT_{iz}}$ and analogously for $\mathcal{M}_x$. The isospin quantum number in the above treatment must respect the GT selection rule, which requires $T_x$ to be within the range of $[T_i-1, T_i+1]$. Otherwise, the corresponding matrix element, $\mathcal{M}_x$ vanishes. 

For simplicity, we further assume that the isospin mixing only occurs with an excited state $\ket{\omega_1T_1T_{fz}}$, whose energy level is closest, among all allowed states, to that of $\ket{\omega_0T_0T_{fz}}$. 
Subsequently, the squared GT matrix element is reduced to  
\begin{equation}\label{eq201}
M_{GT}^2 = \mathcal{M}_0^2 + 2\alpha_1\sqrt{1-\alpha_1^2}\mathcal{M}_0\mathcal{M}_1 + \left( \mathcal{M}_1^2-\mathcal{M}_0^2 \right) \alpha_1^2.  
\end{equation}
Comparing Eq.~\eqref{eq201} to the standard definition of (without the radial mismatch contribution) $M_{GT}^2 = \mathcal{M}_0^2(1-\delta_{C1}^A)$, we obtain $\delta_{C1}^A$ as, 
\begin{equation}\label{CC1}
\delta_{C1}^A = \displaystyle -2\eta\alpha_1 + \left( 1-\eta^2 \right) \alpha_1^2 + \eta\alpha_1^3 + \frac{1}{4}\eta\alpha_1^5 + \mathcal{O}(\alpha_1^7), 
\end{equation}
with $\eta=\mathcal{M}_1/\mathcal{M}_0$. 
A prerequisite for Eq.~\eqref{CC1} is that $\alpha_1$ must be real (a requirement arising from the existence of odd powers of $\alpha_1$). Applying this simplified model to Fermi transitions yields a similar formula for $\delta_{C1}^{V}$ but with $\eta=0$, as Fermi transitions to nonanalog states are isospin forbidden. Generally, one can expect $|\delta_{C1}^A|>|\delta_{C1}^V|$ due to the presence of the first-order term, which results from the relaxation of the GT selection rule. Additionally, the magnitude of $\delta_{C1}^A$ can be greatly enhanced by $\eta$, even though it is an isospin-symmetry quantity. Furthermore, $\delta_{C1}^A$ can be negative, especially if both $\eta$ and $\alpha_1$ are simultaneously positive. On the contrary, $\delta_{C1}^V$ (for Fermi transitions) is always positive since $\delta_{C1}^V=\alpha_1^2$. A negative value of $\delta_{C1}^A$ indicates that isospin mixing leads to a reduction in the absolute GT matrix element. This predicted properties agree remarkably well with the results of exact shell model calculations, as can be observed in Table~\ref{tb:2lev}. 

At the second-order approximation, Eq.~\eqref{CC1} can be solved analytically for the isospin-mixing amplitude $\alpha_1$. The results can be written as:
\begin{equation}\label{alpha}
\alpha_1 = \kappa\left( 1 \pm \sqrt{ 1 + \frac{\delta_{C1}^A}{\eta \kappa} } \right),
\end{equation}
where $\kappa=\eta/(1-\eta^2)$. 
We have numerically verified that only the lower ($-$) sign in Eq.~\eqref{alpha} provides a physical solution for $\alpha_1$. The upper ($+$) sign yields $\alpha_1^2>1$, which violates the conservation of probability. A further study within this simplified model will be given separately in an upcoming paper. 

The two-level mixing model plays an important role in improving the calculated values of $\delta_{C1}^V$ for the superallowed $0^+\rightarrow0^+$ Fermi transition of isotriplets. By utilizing the relation $\delta_{C1}^V=\alpha_1^2$, the calculated values of $\delta_{C1}^V$ can be improved by scaling with the experimental data on the energy separation between the admixed states, $E_1-E_0$. The authors of Ref.~\cite{ToHa2008} suggested taking the average between the scaled and unscaled values of $\delta_{C1}^V$, while treating their difference as a statistical uncertainty. This empirical refinement of $\delta_{C1}^V$ has led to better agreement with the CVC hypothesis. 
Unfortunately, $\delta_{C1}^A$ given in Eq.~\eqref{CC1} contains a first-order term and an additional second-order term in $\alpha_1$ due to the presence of $\mathcal{M}_1$ or $\eta$. This is because the GT process has a more relaxed selection rule for isospin. In certain cases, a direct GT transition to $\ket{\omega_1T_1T_{fz}}$ is energetically forbidden, so the appearance of $\mathcal{M}_1$ in Eq.~\eqref{CC1} is virtually induced by isospin mixing in the final nucleus states. It is also evident that $\delta_{C1}^A$ can be enhanced by $\eta$ even if the isospin-mixing amplitude $\alpha_1$ is small. Our results for $\mathcal{M}_1$ obtained from shell model calculations are listed in Table~\ref{tb:2lev}. 

\begin{table*}[ht!]
\caption{Numerical results for $\delta_{C1}^A$ obtained from large-scale shell-model diagonalizations for a selected model space and effective interaction. The table includes the isospin-symmetry transition matrix elements for the two lowest admixed states of the daughter nuclei, their energy separation, and the parameters of the two-level mixing model, namely $\alpha_1$ and $\eta$. The unit of $\delta_{C1}^A$ 
is in \%. The energy separation, $E_1-E_0$ is in MeV. $T_0$ and $T_1$ are isospin quantum numbers of the lower and the upper admixed states, respectively. 
}\label{tb:2lev}
\begin{threeparttable}
\begin{ruledtabular}
\begin{tabular}{c|c|c|c|c|c|c|c|c|c}
Transition	&	$J_i^\pi	T_i$	&	$J_f^\pi	T_0T_1$	&	Interactions	&	$\mathcal{M}_0$	&	$\mathcal{M}_1$	&	$\eta$	&	$E_1-E_0$	&	$\delta_{C1}^A$	&	$\alpha_1$	\\ 
\hline	
$^8$Li$(\beta^-)^8$Be	&	$2^+	1$	&	$2_1^+	00$	&	CKP/CD	&	0.580	&	3.599	&	6.206	&	9.051	&	-10.775	&	0.008	\\ 
$^8$B($\beta^+$)$^8$Be	&	$2^+	1$	&	$2_1^+	00$	&	CKP/CD	&	0.580	&	3.599	&	6.206	&	9.051	&	-7.597	&	0.006	\\ 
$^9$Li($\beta^-$)$^9$Be	&	$\frac{3}{2}^-	\frac{3}{2}$	&	$\frac{3}{2}_1^-	\frac{1}{2}\frac{1}{2}$	&	CKP/CD	&	0.604	&	0.586	&	0.970	&	4.665	&	-4.785	&	0.025	\\ 
$^9$C($\beta^+$)$^9$B	&	$\frac{3}{2}^-	\frac{3}{2}$	&	$\frac{3}{2}_1^-	\frac{1}{2}\frac{1}{2}$	&	CKP/CD	&	0.604	&	0.586	&	0.970	&	4.665	&	-6.082	&	0.031	\\ 
$^{12}$B($\beta^-$)$^{12}$C	&	$1^+	1$	&	$0_1^+	00$	&	CKP/CD	&	0.960	&	1.861	&	1.938	&	13.467	&	-5.602	&	0.014	\\ 
$^{12}$N($\beta^+$)$^{12}$C	&	$1^+	1$	&	$0_1^+	00$	&	CKP/CD	&	0.960	&	1.861	&	1.938	&	13.467	&	-4.926	&	0.013	\\ 
$^{13}$B($\beta^-$)$^{13}$C	&	$\frac{3}{2}^-	\frac{3}{2}$	&	$\frac{1}{2}_1^-	\frac{1}{2}\frac{1}{2}$	&	CKP/CD	&	1.376	&	0.462	&	0.336	&	8.782	&	-3.589	&	0.058	\\ 
$^{13}$O($\beta^+$)$^{13}$N	&	$\frac{3}{2}^-	\frac{3}{2}$	&	$\frac{1}{2}_1^-	\frac{1}{2}\frac{1}{2}$	&	CKP/CD	&	1.376	&	0.462	&	0.336	&	8.782	&	-1.489	&	0.023	\\ 
$^{17}$N($\beta^-$)$^{17}$O	&	$\frac{1}{2}^-	\frac{3}{2}$	&	$\frac{3}{2}_1^-	\frac{1}{2}\frac{1}{2}$	&	REWIL/CD	&	0.553	&	1.952	&	3.530	&	1.189	&	13.602	&	0.020	\\ 
$^{17}$Ne($\beta^+$)$^{17}$F	&	$\frac{1}{2}^-	\frac{3}{2}$	&	$\frac{3}{2}_1^-	\frac{1}{2}\frac{1}{2}$	&	REWIL/CD	&	0.553	&	1.952	&	3.530	&	1.189	&	49.431	&	0.081	\\ 
$^{20}$F($\beta^-$)$^{20}$Ne	&	$2^+	1$	&	$2_1^+	00$	&	REWIL/CD	&	0.882	&	1.029	&	1.166	&	6.242	&	6.156	&	0.027	\\ 
$^{20}$Na($\beta^+$)$^{20}$Ne	&	$2^+	1$	&	$2_1^+	00$	&	REWIL/CD	&	0.882	&	1.029	&	1.166	&	6.242	&	12.516	&	0.054	\\ 
$^{20}$O($\beta^-$)$^{20}$F	&	$0^+	2$	&	$1_1^+	11$	&	REWIL/CD	&	1.370	&	1.017	&	0.743	&	2.1	&	9.079	&	0.060	\\ 
$^{20}$Mg($\beta^+$)$^{20}$Na	&	$0^+	2$	&	$1_1^+	11$	&	REWIL/CD	&	1.370	&	1.017	&	0.743	&	2.1	&	11.32	&	0.075	\\ 
$^{21}$F($\beta^-$)$^{21}$Ne	&	$\frac{5}{2}^+	\frac{3}{2}$	&	$\frac{3}{2}_1^+	\frac{1}{2}\frac{1}{2}$	&	REWIL/CD	&	0.452	&	1.058	&	2.340	&	4.657	&	5.967	&	0.013	\\ 
$^{21}$Mg($\beta^+$)$^{21}$Na	&	$\frac{5}{2}^+	\frac{3}{2}$	&	$\frac{3}{2}_1^+	\frac{1}{2}\frac{1}{2}$	&	REWIL/CD	&	0.452	&	1.058	&	2.340	&	4.657	&	8.775	&	0.019	\\ 
$^{22}$O($\beta^-$)$^{22}$F &	$0^+	3$	&	$1_1^+	22$	&	REWIL/CD	&	0.245	&	1.878	&	7.663	&	0.673	&	164.147	&	N/A\tnote{*}	\\ 
$^{22}$Si($\beta^+$)$^{22}$Al &	$0^+	3$	&	$1_1^+	22$	&	REWIL/CD	&	0.245	&	1.878	&	7.663	&	0.673	&	127.780	&	N/A\tnote{*}	\\ 
$^{24}$Ne($\beta^-$)$^{24}$Na	&	$0^+	2$	&	$1_1^+	11$	&	REWIL/CD	&	0.877	&	1.513	&	1.726	&	1.804	&	0.694	&	0.002	\\ 
$^{24}$Si($\beta^+$)$^{24}$Al	&	$0^+	2$	&	$1_1^+	11$	&	REWIL/CD	&	0.877	&	1.513	&	1.726	&	1.804	&	19.496	&	0.058	\\ 
$^{25}$Na($\beta^-$)$^{25}$Mg	&	$\frac{5}{2}^+	\frac{3}{2}$	&	$\frac{5}{2}_1^+	\frac{1}{2}\frac{1}{2}$	&	USD/CD	&	0.443	&	0.162	&	0.365	&	1.995	&	1.973	&	0.026	\\ 
$^{25}$Si($\beta^+$)$^{25}$Al	&	$\frac{5}{2}^+	\frac{3}{2}$	&	$\frac{5}{2}_1^+	\frac{1}{2}\frac{1}{2}$	&	USD/CD	&	0.443	&	0.162	&	0.365	&	1.995	&	0.937	&	0.013	\\ 
$^{26}$Na($\beta^-$)$^{26}$Mg	&	$3^+	2$	&	$2_1^+	11$	&	USD/CD	&	0.923	&	0.068	&	0.073   &	1.224	&	0.403	&	0.024	\\ 
$^{26}$P($\beta^+$)$^{26}$Si	&	$3^+	2$	&	$2_1^+	11$	&	USD/CD	&	0.923	&	0.068	&	0.073	&	1.224	&	3.015	&	0.115	\\ 
$^{28}$Al($\beta^-$)$^{28}$Si	&	$3^+	1$	&	$2_1^+	00$	&	USD/CD	&	0.750	&	0.055	&	0.074	&	5.592	&	0.547	&	0.031	\\ 
$^{28}$P($\beta^+$)$^{28}$Si	&	$3^+	1$	&	$2_1^+	00$	&	USD/CD	&	0.750	&	0.055	&	0.074	&	5.592	&	0.621	&	0.034	\\ 
$^{31}$Al($\beta^-$)$^{31}$Si	&	$\frac{5}{2}^+	\frac{5}{2}$	&	$\frac{3}{2}_1^+	\frac{3}{2}\frac{3}{2}$	&	USD/CD	&	0.895	&	1.181	&	1.319	&	2.252	&	-1.77	&	0.007	\\ 
$^{31}$Ar($\beta^+$)$^{31}$Cl	&	$\frac{5}{2}^+	\frac{5}{2}$	&	$\frac{3}{2}_1^+	\frac{3}{2}\frac{3}{2}$	&	USD/CD	&	0.895	&	1.181	&	1.319	&	2.252	&	7.186	&	0.027	\\ 
$^{35}$S($\beta^-$)$^{35}$Cl	&	$\frac{3}{2}^+	\frac{3}{2}$	&	$\frac{3}{2}_1^+	\frac{1}{2}\frac{1}{2}$	&	USD/CD	&	0.534	&	0.060	&	0.112	&	2.625	&	0.052	&	0.002	\\ 
$^{35}$K($\beta^+$)$^{35}$Ar	&	$\frac{3}{2}^+	\frac{3}{2}$	&	$\frac{3}{2}_1^+	\frac{1}{2}\frac{1}{2}$	&	USD/CD	&	0.534	&	0.060	&	0.112	&	2.625	&	2.045	&	0.070	\\ 
$^{35}$P($\beta^-$)$^{35}$S	&	$\frac{1}{2}^+	\frac{5}{2}$	&	$\frac{1}{2}_1^+	\frac{3}{2}\frac{3}{2}$	&	USD/CD	&	0.983	&	0.831	&	0.845	&	2.89	&	-3.322	&	0.020	\\ 
$^{35}$Ca($\beta^+$)$^{35}$K	&	$\frac{1}{2}^+	\frac{5}{2}$	&	$\frac{1}{2}_1^+	\frac{3}{2}\frac{3}{2}$	&	USD/CD	&	0.983	&	0.831	&	0.845	&	2.89	&	2.755	&	0.016	\\ 
\end{tabular}
\end{ruledtabular}
\begin{tablenotes}
{ \raggedright
\item[*] Equation \eqref{alpha} is not valid if $\delta_{C1}^A>1$ since the radial mismatch effect is not included in the two-level mixing model. Nonetheless, as long as the total correction fulfills $\delta_C^A\le1$, these values of $\delta_{C1}^A$ remain physically allowable. \\
}
\end{tablenotes}
\end{threeparttable}
\end{table*}

\subsection{Exact shell-model calculation of $\delta_{C1}^A$} 

Following the standard shell-model procedure, the many-particle wave function of the initial and final states can be obtained by diagonalizing the Hamiltonian matrix in a harmonic oscillator basis within a specified valence space. The effective shell-model Hamiltonian, denoted as $H$, is generally decomposed of two terms:
\begin{equation}
H = H_0 + V_{INC},
\end{equation}
with $[H_0,T^2]=0$ and $[V_{INC},T^2]\ne 0$. Within the phenomenological approach~\cite{USD,USDab,gx1a,kb3g,jun45,Cohen1965,ZBM1968}, $H_0$ includes isoscalar single-particle energies, typically taken as the average of the proton and neutron energies of closed-shell plus or minus one nucleon nuclei. The two-body matrix elements of $H_0$ are obtained through a least-square fitting process to reproduce the binding and excitation energies of multiple nuclei within the valence space. The isospin non-conserving component $V_{INC}$ is then added on top of the isoscalar effective Hamiltonian $H_0$. According to Ormand and Brown~\cite{OrBr1985,OrBr1989,OrBr1989x,OrBr1995,Lam2013}, the two-body part of $V_{INC}$ comprises the Coulomb interaction and a sum of Yukawa potentials representing nuclear CIB (charge-independent-breaking) and CSB (charge-symmetry-breaking) forces, with their strength constrained by the experimental data on the $b$ and $c$ coefficients of the isobaric multiplet mass equation. The matrix elements of the one-body part of $V_{INC}$ can be directly replaced with experimental data on the isovector single-particle energies of the specified valence space. 

In general, phenomenological effective interactions exhibit higher precision compared to those derived from realistic nucleon-nucleon potentials or from chiral effective potentials. 
The calculation of $\delta_{C1}^A$, however, can still be affected by errors due to the dependence on the type and number of experimental data used to constrain the interaction matrix elements, as well as some fine details of the fitting protocol. Therefore, additional refinements are often necessary to mitigate such issues, particularly for the superallowed $0^+\rightarrow 0^+$ Fermi transition of isotriplets and the mirror $\beta$ decay of isodoublets, thanks to significant experimental advancements. 
For instance, Lam {\it et al.}\cite{Lam2013} found that the interaction precision can be slightly improved by using harmonic oscillator parameter values extracted case-by-case from the measured charge-radius data, instead of relying on global parameterizations\,\cite{Kir2006,BLOMQVIST1968545}. On the other hand, Towner and Hardy~\cite{ToHa2008} employed a simpler schematic interaction for the nuclear component of $V_{INC}$ and locally refitted its matrix elements for each mass multiplet using the experimental data on the $b$ and $c$ coefficients as constraints. Furthermore, the results of the two-level mixing model discussed in the previous subsection, namely $\delta_{C1}^V\propto \alpha_1^2$, were used to improve the calculated $\delta_{C1}^V$ values for Fermi transitions by scaling them with the measured energy separation between the admixed levels in daughter nuclei, as mentioned earlier. Unfortunately, this scaling technique does not seem to be applicable for GT transitions due to the presence of the LO and higher-order terms in $\alpha_1$ in Eq.~\eqref{alpha}. In general, $\delta_{C1}^A$ for GT transitions is expected to be stronger because the operator $\bm{\sigma}\bm{\tau}_\pm$ has a more relaxed selection rule. 

In this work, we do not introduce any new developments in the isospin nonconserving effective interaction. Instead, we employ existing effective Hamiltonians for $H_0$ in our numerical calculation of $\delta_{C1}^A$. For light nuclei with mass number between 7 and 12, we use the CKPOT interaction of Cohen and Korath~\cite{Cohen1965,Cohen1967} along with some other available interactions in the same valence space, including those from Ref.~\cite{VANHEES198861,s.afr}. For nuclei with mass number $13 \le A \le 24$, we select the interaction of McGrory-Wildenthal and Reehal~\cite{PhysRevC.7.974}. For heavier nuclei up to $A=35$, we utilize the universal SD interaction of Chung and Wildenthal~\cite{USD} as well as the two new versions of Brown and Richter~\cite{USDab}. The diagonalization of the Hamiltonian matrix is performed without truncation in the full model spaces using the code NuShellX@MSU~\cite{NuShellX}. The isospin nonconserving part $V_{INC}$ is taken from the works of Ormand and Brown~\cite{OrBr1985,OrBr1989,OrBr1989x,OrBr1995} without further refinement. We reexamine the pure GT transitions studied in Ref.\,\cite{Smirnova2003} within the generalized formalism. We also consider $^{24}$Si and $^{24}$Ne, for which the $ft$ values and mirror asymmetry have been available with good precision since 2009~\cite{PhysRevC.80.044302}. Furthermore, the recent measurements of $^{26}$P and $^{26}$Na are investigated~\cite{sym13122278}. 

Our numerical results are given in Table~\ref{resC1}. It is evident that $\delta_{C1}^A$ exhibits significant variability from transition to transition,
and negative values are obtained in many cases. Therefore, the absolute GT transition matrix element could either be enhanced or quenched due to the ISB effect, depending mainly on the parameters $\eta$ and $\alpha_1$ as seen from the two-level mixing model discussed in the previous subsection. As an obvious example, the $\delta_{C1}^A$ value obtained for $^{17}$Ne($\beta^+$)$^{17}$F is almost 50\,\%, 
because $\eta$ is quite large, and at the same time, the energy separation between the admixed states, $E_1-E_0$, is quite small. On the contrary, $\delta_{C1}^V$ for Fermi transitions is always positive~\cite{HaTo2020,Bhattacharya}, meaning that the exact value of the absolute Fermi matrix element is always smaller than its isospin symmetry limit. It is interesting to remark that all these behaviors can be qualitatively explained by the two-level mixing model. 

We emphasize that the transitions $^{24}$Ne($\beta^-$)$^{24}$Na and $^{24}$Si($\beta^+$)$^{24}$Al have been studied within the first-principle shell model in Ref.~\cite{PhysRevC.80.044302}. In that study, it was discovered that the observed strong mirror asymmetry in the GT strengths is mainly due to the Thomas-Erhman shift~\cite{PhysRev.88.1109,PhysRev.81.412}, 
and any attempt to describe it without modifying the proton $2s_{1/2}$ energy would be unsuccessful. Surprisingly, the experimental data in question are also well reproduced by our present calculations, which utilize the effective interaction from Ref.~\cite{PhysRevC.7.974} that includes the $1p_{1/2}$ orbital. 

We note that our adopted values of $\delta_{C1}^A$ are taken as the average of those obtained with different effective interactions. 
However, for nuclei with $17\le A\le 24$, only the interaction from Ref.~\cite{PhysRevC.7.974} is considered. The other available interactions in the same model space, for example those used in Ref.~\cite{PhysRevLett.21.39}, do not seem to provide a reasonable result. The present study serves as a proof-of-concept, and there are several details and possible improvements that could be addressed in a separate paper. For instance, further investigation on scaling $\delta_{C1}^A$ with the energy separation between the two admixed levels may still be reasonable, especially when $\delta_{C1}^A$ is small enough to neglect the higher-order term in Eq.~\eqref{CC1}. 


\section{Radial mismatch correction}\label{radial}
\subsection{General formalism}

With the long-range Coulomb interaction as the main driver, the isospin admixtures in the initial and final nuclear states of $\beta$ decays can extend further beyond the modest shell-model configuration spaces. Consequently, the calculation of $\delta_{C1}^A$ using first-principle shell models is generally underestimated. To account for the couplings to states lying outside the model spaces, the correction $\delta_{C2}^A$ due to the difference between proton and neutron radial wave functions must be considered, 
\begin{equation}\label{qC2} 
\displaystyle \delta_{C2}^A = \frac{2}{\mathcal{M}_{GT}} \sum_{k_ak_b\pi} \theta_{ab}^{\lambda} \Theta_{abfi}^{\pi\lambda} \Lambda_{ab}^{\pi} \xi_{ab} A^T(f;\pi a)A^T(i;\pi b), 
\end{equation}
where the label $\pi$ refers to the virtual intermediate $(A-1)$-nucleon states, and the labels $a$ and $b$ correspond to the single-particle states, specifically $i=k_im_i\tau_i=n_il_ij_im_i\tau_i$ with $i$ being either $a$ or $b$. The quantities $A^T(f;\pi a)$ and $A^T(i;\pi b)$ represent the spectroscopic amplitudes for removing one nucleon from single-particle states $a$ and $b$ adjacent to the final and initial states, respectively. 
These amplitudes are defined as follows: 
\begin{equation}\label{af}
A^T(f;\pi a) =\frac{ (f|| a_a^\dagger ||\pi) }{ \sqrt{2J_f+1} }, 
\end{equation}
and 
\begin{equation}\label{ai}
A^T(i;\pi b) =\frac{ (i|| a_b^\dagger ||\pi) }{ \sqrt{2J_i+1} }, 
\end{equation}
where the double bars in the reduced matrix elements denote a reduction in angular momenta. 
It should be noted that we use round brackets for many-particle states with definite isospin. 
These amplitudes can be obtained using a well-established isospin-invariant effective Hamiltonian such as those listed in the column `Interactions' of Table~\ref{tb:2lev} and the harmonic oscillator basis. 

\begin{table*}[ht!]
\caption{Analytical expressions of the function $\Theta_{abfi}^{\pi\lambda}$. The shorthand notations of $\hat{X}=X(X+1)$ and $X_{ijk...}^{lmn...}=X_i+X_j+X_k+...-X_l-X_m-X_n-...$ are used for the angular momenta in these expressions.}\label{tb:Theta}
\begin{ruledtabular}
\begin{tabular}{c|c|c|c}
\backslashbox{$j_a$}{$J_f$} & $J_i+1$ & $J_i$ & $J_i-1$ \\
\hline
$j_b+1$ & $\displaystyle-\left[ \frac{J_{f b}^\pi(J_{\pi i b}+2)(J_{\pi i b}+3)(J_{i b}^\pi+2)}{4j_a(2j_b+1)(2j_b+3)(2J_i+3)}\right]^\frac{1}{2}$ & $\displaystyle-\frac{J_{ia}^\pi}{2}\left[ \frac{(2J_i+1)(J_{\pi i b}+2)J_{\pi i}^b}{j_a(2j_b+1)(2j_b+3)\hat{J}_i} \right]^\frac{1}{2}$ & $\displaystyle-\left[ \frac{J_{a\pi}^iJ_{a\pi}^fJ_{i\pi}^aJ_{i\pi}^b}{4j_a(2j_b+1)(2j_b+3)J_i} \right]^\frac{1}{2}$  \\
\hline
$j_b$ & $\displaystyle-\left[ \frac{J_{i b}^i(J_{\pi i b}+2)(J_{\pi i}^b+1)(J_{i b}^\pi+1)}{4\hat{j}_b(2j_b+1)(2J_i+3)} \right]^\frac{1}{2}$ & $\displaystyle\left[ \frac{(2J_i+1)[-\hat{J}_\pi+\hat{j}_b+\hat{J}_i]}{4\hat{j}_b(2j_b+1)\hat{J}_i} \right]^\frac{1}{2} $  & $\displaystyle-\left[ \frac{(J_{ib\pi}+1)J_{b\pi}^fJ_{i\pi}^bJ_{bi}^\pi}{4\hat{j}_b(2j_b+1)J_i} \right]^\frac{1}{2}$ \\
\hline 
$j_b-1$ & $\displaystyle-\left[ \frac{J_{\pi b}^fJ_{\pi b}^iJ_{\pi i}^a(J_{\pi i}^b+2)}{4(2j_b-1)j_b(2j_b+1)(2J_i+3)} \right]^\frac{1}{2}$ & $\displaystyle\left[ \frac{(2J_i+1)(J_{ib\pi}+1)J_{b\pi}^i(J_{i\pi}^b+1)J_{bi}^\pi}{4(2j_b-1)j_b(2j_b+1)\hat{J}_i} \right]^\frac{1}{2}$  & $\displaystyle-\left[ \frac{J_{ib\pi}(J_{ib\pi}+1)J_{ai}^\pi J_{bi}^\pi}{4(2j_b-1)j_b(2j_b+1)J_i} \right]^\frac{1}{2}$ \\
\end{tabular}
\end{ruledtabular}
\end{table*}

The factor $\Lambda_{ab}^{\pi}$ in Eq.~\eqref{qC2} denotes the deviation from unity of the overlap integral between proton and neutron radial wave functions. It is defined as: 
\begin{equation}\label{over}
\Lambda_{ab}^{\pi} = 1-\int_0^\infty R_a^{\pi}(r)R_b^{\pi}(r) r^2 dr >0, 
\end{equation}
here the label $\pi$ indicates that $\Lambda_{ab}^{\pi}$ is evaluated with radial wave functions whose asymptotic form matches the separation energies relative to a specific excited state $\ket{\pi}$ of the intermediate $(A-1)$-nucleon system. For example, if $\ket{\pi}$ is the ground state, then the proton and neutron separation energies are $S_p$ and $S_n$, respectively, which are well-known from experiment and can be found in any atomic mass table~\cite{AME2012}. If, however, $\ket{\pi}$ is an excited state with excitation energy $E_x^\pi$, then the proton and neutron separation energies are $S_p+E_x^\pi$ and $S_n+E_x^\pi$, respectively. This important constraint for radial wave functions can be implemented by refitting, for example, the depth of the mean field potential separately for protons and neutrons. Note that $\Lambda_{ab}^{\pi}=0$ if one employs an isospin-symmetry single-particle basis such as the harmonic oscillator basis. 

We recall that the introduction of intermediate state dependence for Fermi transitions has resulted in a systematic increase in $\delta_{C2}^V$ (about\,20~\%). This improvement leads to much better agreement with the predictions of the Standard Model compared to the results obtained using the closer approximation. 

The function $\Theta_{abfi}^{\pi\lambda}$ appears in Eq.~\eqref{qC2} due to the conversion of the one-body transition densities (OBTD$^T$) into the products of spectroscopic amplitudes for protons and neutrons by introducing a complete set of intermediate states of the $(A-1)$-nucleon system. It is given by:
\begin{equation*}\label{dd}
\begin{array}{ll}
\Theta_{abfi}^{\pi\lambda} & = \displaystyle (-1)^{J_f+J_\pi+j_a+\lambda}\sqrt{(2J_i+1)(2J_f+1)} \\[0.18in]
& \displaystyle \times\left\{
\begin{array}{lll}
J_i & J_f & \lambda \\[0.18in]
j_b & j_a & J_\pi
\end{array}
\right\}, 
\end{array}
\end{equation*}
where the intermediate state spin $J_\pi$ is determined by the $6j$ symbol. Analytical expressions for $\Theta_{abfi}^{\pi\lambda}$ are given in Table~\ref{tb:Theta}. 
For $J_f>J_i$ ($J_f=J_i+1$), the product $\theta_{ab}^\lambda \Theta_{abfi}^{\pi\lambda}$ is negative for a GT transition between spin-up states or $j_b=j_a=l+\frac{1}{2}$ and positive when $j_a=l-\frac{1}{2}$ irrespective of $j_b$. In the case of $J_f\le J_i$, this product is negative for a GT transition from a spin-down state to the corresponding spin-up state and positive for the other configurations of $j_b$ and $j_a$. This property could lead to an important cancellation in $\delta_{C2}^A$, in particular when the sign of the product $A^T(f;\pi a)A^T(i;\pi b)$ is consistent (either the same or opposite to) with that of $\theta_{ab}^\lambda \Theta_{abfi}^{\pi\lambda}$. Therefore, special attention must be paid to the selection of configuration space; for example, the model space should always cover both spin-up and spin-down states. However, this picture could be disturbed by the interplay of $\Lambda_{ab}^{\pi}$, which tends to vanish for the deeper bound spin-up state of a given model space or by $A^T(f;\pi a)A^T(i;\pi b)$ when its sign is inconsistent with that of $\theta_{ab}^\lambda \Theta_{abfi}^{\pi\lambda}$. A further discussion on the structural properties of $\delta_{C2}^A$ is given in Section~\ref{iso}. 

\subsection{Realistic Woods-Saxon potential}

The comparative study of the radial mismatch correction between Woods-Saxon and Skyrme-Hartree-Fock radial wave functions was first carried out by Ormand and Brown~\cite{OrBr1985}. This pioneering study revealed that the correction values for several superallowed $0^+\rightarrow 0^+$ Fermi transitions of isotriplets obtained with Skyrme-Hartree-Fock were consistently lower than those obtained with Woods-Saxon wave functions. This discrepancy has been the subject of intense debate over the last few decades. Recently, Xayavong and Smirnova~\cite{XaNa2022} pointed out several deficiencies in the Hartree-Fock mean field, including finite size effects, the Slater approximation to the Coulomb exchange, the absence of charge-symmetry breaking forces, spurious isospin mixing, and the Wigner energy effect. Moreover, the Hartree-Fock mean field is not suitable for our method as it requires that the radial wave functions should match the experimental separation energies and charge radii, which, for the Woods-Saxon potential, can be achieved by refitting the depth and length parameter. 
For those reasons, we will consider only the phenomenological Woods-Saxon mean field for the present study. Following Ref.~\cite{XaNa2018}, the Woods-Saxon potential is expressed as:
\begin{equation}\label{ws}
\begin{array}{ll}
V(r) =& \displaystyle V_0f_0(r) - V_s \left(\frac{r_s}{\hbar}\right)^2\frac{1}{r}\frac{d}{dr}f_s(r)\braket{\bm{l}\cdot\bm{\sigma}} \\[0.15in]
&\displaystyle+ V_{\text{surf}}(r) + \left(\frac{1}{2}-t_z\right)V_C(r),
\end{array}
\end{equation}
where the functions $f_i(r)$ are defined as:
\begin{equation}\label{fi}
\displaystyle f_i(r) = \frac{1}{1+\exp\left( \frac{r-R_i}{a_i} \right)},
\end{equation}
with $i=0$ or $s$ denoting either the central or spin-orbit terms. The assumption of $a_0=a_s$ is usually adopted due to the lack of experimental constraints. However, an alternative suggestion proposes a smaller spin-orbit radius $(R_s< R_0)$ because the two-body spin-orbit interaction has a shorter range~\cite{BohrMott}. For instance, the Seminole parameterization given in Ref.~\cite{SWV} yields $R_s/R_0=0.921$. 
The expectation value $\braket{\bm{l}\cdot\bm{\sigma}}$ is given by
\begin{equation}
\braket{\bm{l}\cdot\bm{\sigma}} = \left\{
\begin{array}{lll}
l & \text{if} & j=l+\frac{1}{2} \\[0.1in]
-(l+1) & \text{if} & j=l-\frac{1}{2}. 
\end{array}
\right.
\end{equation}
The symmetry term considered in Ref.~\cite{XaNa2018} is omitted because it is redundant with the central term and it can be fully recovered by readjusting $V_0$ for protons and neutrons separately to reproduce their respective separation energies at the ground state of the intermediate nucleus. 

Following the usual practice, the Coulomb repulsion is accounted for using the approximation of a uniformly charge distributed sphere:
\begin{equation}\label{eq1}
V_C(r) = (Z-1)e^2 \left\{
\begin{array}{ll}
\displaystyle \frac{1}{r}, & r>R_C \\[0.1in]
\displaystyle \frac{1}{R_C}\left( \frac{3}{2} - \frac{r^2}{2R_C^2} \right), & \text{otherwise,}
\end{array}
\right.
\end{equation}
where $R_C$ is usually taken as $R_C=r_0(A-1)^\frac{1}{3}$ with $r_0\approx 1.26$~fm~\cite{SWV}, alternatively it can be extracted from the charge radius $R_{ch}$ via~\cite{Elton}
\begin{equation}\label{eq2}
\displaystyle R_C^2 = \frac{5}{3} R_{ch}^2 - \frac{5}{2} \sum_{i=1}^3 \theta_i r_i^2 - \frac{5}{4}\left( \frac{\hbar}{mc} \right)^2 + \frac{5}{2} \frac{b^2}{A}. 
\end{equation}
The nuclear oscillator length parameter is given by $b^2\approx A^\frac{1}{3}$~fm$^2$, while an improved parameterization of $b^2 $ can be found in Ref.~\cite{Kir2006}. 
The last three terms on the right-hand-side of Eq.~\eqref{eq2} account for the internal structure of proton, where $\sum_i \theta_i r_i^2=0.518$~fm$^2$~\cite{XaNa2022}, the Darwin-Foldy term ($\hbar/mc=0.21$~fm) and the center-of-mass (COM) motion, respectively. 

In order to preserve the fundamental symmetries, the phenomenological effective potential, such as Woods-Saxon, is typically treated as a nuclear mean field created by the core of $(A-1)$ nucleons. Excluding the contribution from the last nucleon also serves as a self-interaction correction due to the missing exchange terms. Thus, the radius parameters are given by $R_i=r_i (A-1)^\frac{1}{3}$. The length parameter of the volume term, $r_0$, is readjusted to reproduce the charge radius when experimental data are available. The other parameters are kept fixed, and their numerical values are taken from Ref.~\cite{XaNa2018}.

The conventional part of the Woods-Saxon potential (volume, spin-orbit, and Coulomb terms) has been optimized through a global fit to reproduce some bulk properties of closed-shell nuclei and the energy of single-particle or single-hole states. Therefore, this conventional part should not be locally readjusted for a given open-shell nucleus. Instead, it is fundamentally more appropriate to add an extended term to compensate for beyond mean-field effects. For the present work, we consider the surface-peaked term $V_{\text{surf}}(r)$, defined below:
\begin{equation}
V_{\text{surf}}(r) = V_s\left( \frac{\hbar}{m_\pi c} \right)^2 \frac{1}{a_s r}\exp\left( \frac{r-R_s}{a_s} \right) [ f_s(r) ]^2,
\end{equation}
where $m_\pi$ is the mass of the pion, and $V_s$ is a free parameter to be adjusted to reproduce excitation energies of the intermediate nucleus, in addition to adjusting $V_0$ at the ground state. The surface-peaked term $V_{\text{surf}}(r)$ plays an important role in the optical model because nuclear reactions take place predominantly around the surface region of the target nucleus~\cite{PINKSTON1965641}. This additional term was also considered in the studies of superallowed $0^+\rightarrow 0^+$ Fermi $\beta$ decay of isotriplets~\cite{ToHa2008,XaNa2018}.  

\subsection{Spin-orbit effect}

The spin-orbit effect on GT transition matrix element has not been pointed out before, and it highlights an additional deficiency of the conventional shell model, which employs the isospin-invariant harmonic-oscillator basis. Addressing this effect could lead to a deeper understanding of nuclear structure and related phenomena. 

Being driven by the axial-vector current, the GT process can also take place between orbits of opposite spin alignment, such as between $1p_{1/2}$ and $1p_{3/2}$ or vice versa. As a result, the corresponding radial mismatch factor, $\Lambda_{ab}^{\pi}$, can be influenced by the radial coordinate dependence of the spin-orbit term of the realistic mean-field potential. Since this effect interferes with the radial mismatch component of the ISB correction, it should be investigated within the present theoretical framework. 


Since the nuclear spin-orbit term is effectively negative~\cite{PhysRev.78.16,PhysRev.78.22}, the single particle potential become deeper 
in the presence of spin-orbit interaction compared to the case without it when the spin is aligned with the orbital angular momentum, making the state more tightly bound. On the contrary, when spin and orbital angular momentum are anti-aligned, the system's energy is higher. In a realistic mean field, such as the self-consistent Hartree-Fock or the phenomenological Woods-Saxon potential, the asymptotic radial wave function is known to exhibit exponential decay, and its slope in logarithmic scale is determined by the energy. Consequently, the radial overlap integral between states that are spin-orbit partners would be slightly smaller than unity even at the isospin-symmetry limit, due to the binding energy effect. Notably, the spin-orbit form factor has a peak near the nuclear surface~\cite{Thomas}, which is the region where the potential significantly affects the properties of a single particle near the Fermi level~\cite{10.1093/ptep/ptab136}. Therefore, the replacement of the harmonic oscillator basis of the shell model with a realistic one corrects not only for the ISB effect but also for the spin-orbit effect on radial wave functions. As a result, a larger $\delta_{C2}$ can be expected for a GT transition. 
\begin{center}
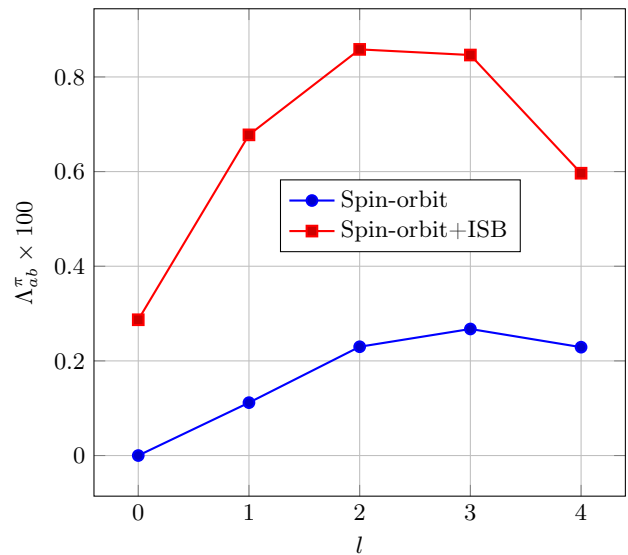
\begin{figure}[ht!]
    \begin{tikzpicture}
     \begin{groupplot}[width=\linewidth, height=0.45\textwidth, group style={group size=1 by 1,xlabels at=edge bottom},
      ylabel=$\Lambda_{ab}^{\pi}\times 100$, xlabel=$l$, grid=both]  
       \nextgroupplot[legend style={at={(0.35,0.65)}, anchor=north west,legend columns=1},legend cell align={left}]
        \addplot+[thick,error bars/.cd,y dir=both,y explicit] table[x=l,y=omega0] {omega.dat}; 
        \addlegendentry{Spin-orbit}
        \addplot+[thick,error bars/.cd,y dir=both,y explicit] table[x=l,y=omega1] {omega.dat}; 
        \addlegendentry{Spin-orbit+ISB}
      \end{groupplot}
    \end{tikzpicture}
\caption{\label{fig1}(Color online) Radial mismatch factor,  $\Lambda_{ab}^{\pi}$ between spin-up (protons) and spin-down (neutrons) states as a function of orbital angular momentum quantum number $l$. For this illustrative example, we consider the Gamow-Teller transition from the ground state of $^{100}$Sn to the first $1^+$ state of $^{100}$In. The intermediate nucleus $^{99}$In is assumed to be in the ground state. The radial wave functions are taken as eigenfunctions of a fixed-parameter Woods-Saxon potential.}
\end{figure}
\end{center}

Fig.~\ref{fig1} displays a rough estimate of the spin-orbit contribution to $\Lambda_{ab}^{\pi}$ as a function of $l$ for the GT transition from the ground state of $^{100}$Sn (with $0^+$ spin-parity) to the first $1^+$ state of $^{100}$In
To ensure minimal ISB contamination, the radial wave functions used in this calculation are obtained by solving a fixed-parameter Woods-Saxon potential with the Coulomb and nuclear symmetry terms switched off. Furthermore, we assume that the intermediate nucleus $^{99}$In is in its ground state. It is essential to note that this example considers only nodeless states ($n=1$); for higher $n$ states, the radial mismatch would be stronger due to the binding-energy effect. 

In general, the spin-orbit contribution to $\Lambda_{ab}^{\pi}$ increases with $l$ (for a given $n$), as the amount of splitting in $\braket{\bm{\sigma}\cdot\bm{l}}$ between spin-up and spin-down states is proportional to $(2l+1)$. This behavior is clearly observed for $l\le 2$, as shown in the numerical results presented in Fig.~\ref{fig1}. However, for higher $l$ values, the factor $\Lambda_{ab}^{\pi}$ tends to decrease due to the centrifugal barrier increasing faster than the spin-orbit splitting. Nonetheless, this illustrative example reveals that the spin-orbit contribution can still be significant, reaching up to 30\,\% of the total $\Lambda_{ab}^{\pi}$, particularly for high $l$ orbits. Conducting a more rigorous study of the spin-orbit contribution presents challenges, as the current theoretical approach requires readjustments of the Woods-Saxon parameters to accurately reproduce available data on separation energies and charge radii. Consequently, it becomes difficult to investigate individual terms of the potential by switching them on and off.


\subsection{Charge radius constraint} 

The charge radii provide a sensitive constraint for the radial wave functions of protons. 
Thus, it is crucial to adjust the length parameter $r_0$ in order to accurately 
replicate the experimental data on charge radii. 
This adjustment ensures the correctness of the radial wave functions of protons in the calculation of $\delta_{C2}$. Since the work of Xayavong~\cite{Xthesis}, the formulation of the root-mean-square (rms) radius of a point-like nucleon distribution has been generalized by considering the fragmentation of single-nucleon strengths and taking into account the excitation energies of intermediate system. The expression is given by:  
\begin{equation}\label{rad}
\displaystyle \braket{\boldsymbol{r}_{\tau_a}^2}=\frac{1}{Z}\sum_{k_a\pi} \mathcal{A}(i;\pi a)^2 \int_0^\infty r^4 |R_a^{\pi}(r)|^2dr. 
\end{equation} 
The wave function normalization implies that $\sum_{\pi} \mathcal{A}(i;\pi a)^2=\braket{n_a}$. In order to convert this expression into charge radii, it must be supplemented with the corrective terms in Eq.~\eqref{eq2}. 
This generalization allows for a self-consistent adjustment of the depth or the strength of the surface-peaked term and length parameters in a two-parameter grid. 
It was found that this generalization yields smaller rms radii, and subsequently larger $r_0$ compared to the traditional approach when fitting to the same experimental data. 

We notice that the rms radii of neutron distribution are unknown for most nuclei, since neutron is practically a neutral particle. In the previous calculations of nuclear structure corrections for the superallowed $0^+\rightarrow 0^+$ Fermi transitions of isotriplets, the value of $r_0$ obtained for protons in mother nuclei was also used for neutrons in daughter nuclei. 
In general, the shape of the mother and of daughter nuclei of Fermi transitions could be completely different, the shape asymmetry in mirror nuclei have been recently reported~\cite{PhysRevLett.126.072501}. This assumption of identical radius could be even more unrealistic in GT transitions, where the initial and final states are generally not isobaric analogue states. Furthermore, the final states of GT transitions are usually an excited or resonance state. Therefore, $\delta_{C2}$ values for GT transitions would generally be less accurate than for Fermi transitions, especially the $\beta^-$ partners. 

In the present study, experimental data on charge radii is unavailable for the majority of considered nuclei. Consequently, in such instances, we have assigned the value of $r_0$ to the standard value.
Despite this limitation, we have observed that for certain cases, the difference between $\delta_{C2}^A$ values obtained with and without the inclusion of the surface-peaked term--treated as a source of uncertainty for $\delta_{C2}^A$--appears to be larger than the fluctuation due to the variation of $r_0$ around its standard value. In such instances, it becomes unnecessary to consider additional constraints on the radial wave functions, such as accurate charge radius data, in addition to the separation energy constraint and the Standard Model filter. Nonetheless, these additional constraints could potentially aid in distinguishing or identifying a specific extended term within the Woods-Saxon potential.
It is desired that a strong motivation is provided by our study for experimental groups worldwide to undertake new measurements of charge radii for the nuclei considered in this research. 

On the other hand, even though the calculations of $\delta_C$ for individual GT transitions suffer from numerous sources of uncertainty, the uncertainty on the mirror asymmetry $\delta$ would be greatly reduced since it is sensitive to the difference between $\delta_C^A$ values for the $\beta^+$ and $\beta^-$ partners. It was pointed out in Ref.~\cite{XaNa2022} that $\delta_{C2}^V$ is very sensitive to the difference between neutron and proton separation energies. One can also expect this behavior of $\delta_{C2}^V$ for GT transitions. 
 
\subsection{Convergence property} 

The decomposition of $\delta_{C2}^A$ in terms of intermediate states enables us to constrain the radial wave functions by fitting the separation energies concerning a specific excitation of the intermediate nucleus. For Fermi transitions, this empirical refinement yields significantly larger $\delta_{C2}^V$ values and improves the agreement with the Standard Model compared to the closer approximation~\cite{ToHa2002,XaNa2018}. However, evaluating $\delta_{C2}^A$ using the refined expression~\eqref{qC2} involves summing over intermediate states of spectroscopic amplitudes weighted with the radial mismatch and some spin-angular factors. Consequently, an exact calculation of $\delta_{C2}^A$ is much more computationally intensive than $\delta_{C1}^A$, which requires only an initial and a final state. However, as an excitation of the $(A-1)$-nucleon system makes single-particle states in the initial and final nuclei more bound, the corresponding radial wave functions become less sensitive to minor terms of the mean-field potential, such as the Coulomb and nuclear symmetry terms, which violate isospin symmetry. Due to this property, the factor $\Lambda_{ab}^{\pi}$ should decrease monotonically as a function of the excitation energy of the $(A-1)$-nucleon system. Additionally, the single-particle strengths (spectroscopic factors) generally have less importance for high-energy intermediate states. Therefore, it is possible to establish a reasonable cut-off at a certain energy level, ensuring that the excluded contribution is below the desired precision. This allows for a more manageable and accurate calculation of $\delta_{C2}^A$ while maintaining the essential physics of the system. 

On the other hand, a traditional approach is to directly introduce the complete set of intermediate states into the transition matrix element without defining the ISB correction~\cite{Smirnova2003}. In this manner, the result, excluding the isospin-mixing contribution, can be expressed as follows: 
\begin{equation}\label{mgt}
\displaystyle M_{GT} = \hat{J}_i\hat{J}_f\sum_{k_ak_b\pi} \theta_{ab}^\pi \Theta_{abfi}^{\pi\lambda} \Omega_{ab}^{\pi} \xi_{ab} A^T(f;\pi a)A^T(i;\pi b), 
\end{equation}
where $\hat{J}$ is the shorthand for $\sqrt{2J+1}$, and $\Omega_{ab}^{\pi}$ is the overlap integral, defined as $\Omega_{ab}^{\pi}=1-\Lambda_{ab}^{\pi}$. It is important to note that the high energy limit of $\Omega_{ab}^{\pi}$ and $\Lambda_{ab}^{\pi}$ is 1 and 0, respectively. For simplicity, we assume that $M_{GT}$ converges linearly, and its rate of convergence is given by: 
\begin{equation}
\displaystyle R_{M_{GT}} = \lim_{E_x^{\pi*}\to\infty}\frac{ \Omega_{ab}^{\pi**} A^T(f;\pi^{**} a)A^T(i;\pi^{**} b) }{ \Omega_{ab}^{\pi*} A^T(f;\pi^* a)A^T(i;\pi^* b) },
\end{equation}
where $\ket{\pi^*}$ represents a high energy excited state at which the sum over the index $\pi$ can be truncated, and $\ket{\pi^{**}}$ denotes the next excited state. The schematic representation of these states is illustrated in Fig.~\ref{lev}. Note that $\pi^*$ could be a high energy resonance state. 

\begin{center}
\begin{figure}[ht!]
    \begin{tikzpicture}[framed]
        \draw[ultra thick,black] (0,0)--(2,0) node[above,midway] {$\vdots$} node[below,midway] {g.s.}; 
        \draw[ultra thick,red] (0,1.7)--(2,1.7) node[above,midway] {$\ket{\pi*}$} node[right] {cut-off level}; 
        \draw[ultra thick,black] (0,2.5)--(2,2.5) node[above,midway] {$\ket{\pi**}$}; 
    \end{tikzpicture}
\caption{Schematic representation of the cut-off level for intermediate states. $\ket{\pi^*}$ is a high-energy state such that $\Omega_{ab}^{\pi*}\approx1$.}
\label{lev}
\end{figure}
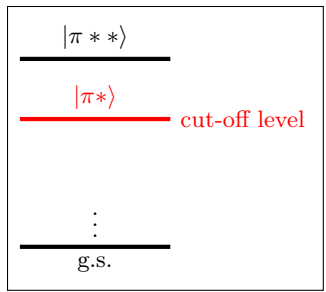
\end{center}

Similarly, the linear rate of convergence of $\delta_{C2}^A$ derived using the expression~\eqref{qC2} is given by: 
\begin{equation}
    R_{\delta_{C2}^A} = \lim_{E_x^{\pi*}\to\infty}\frac{ \Lambda_{ab}^{\pi**} A^T(f;\pi^{**} a)A^T(i;\pi^{**} b) }{ \Lambda_{ab}^{\pi*} A^T(f;\pi^* a)A^T(i;\pi^* b) }.  
\end{equation}

An interesting feature is that the ratio of these rates does not depend on the spectroscopic amplitudes. In other words:
\begin{equation}\label{rat}
\frac{R_{\delta_{C2}^A}}{R_{M_{GT}}} = \lim_{E_x^{\pi*}\to\infty} 
\left( \frac{ 1 -\Omega_{ab}^{\pi**} }{\Omega_{ab}^{\pi**} } \right)
\left( \frac{ \Omega_{ab}^{\pi*} }{1-\Omega_{ab}^{\pi*} } \right) <1 ,
\end{equation}
where $ \Omega_{ab}^{\pi*} < \Omega_{ab}^{\pi**}$ due to the binding-energy effect, with $\ket{\pi^{**}}$ lying above $\ket{\pi^*}$. The expression~\eqref{rat} indicates that $R_{\delta_{C2}^A}<R_{M_{GT}}$, signifying that $\delta_{C2}^A$ converges faster than $M_{GT}$. In fact, the convergence speed of $M_{GT}$ is determined only by the single-particle strengths or spectroscopic factors, which should be normalized to the occupation numbers of valence orbits. This property has been numerically studied for the superallowed $0^+\rightarrow 0^+$ Fermi transition of isotriplets~\cite{XaNa2018}, and was one of the main reasons for the introduction of $\delta_C^V$. Importantly, Eq.~\eqref{rat} does not rely on the structure of the initial and final states, or the selection rule of the transition operator, making it applicable to any type of $\beta$ transitions. 

The properties of $\Lambda_{ab}^\pi$ discussed above should also be applicable to GT transitions. However, since the spectroscopic amplitudes in Eq.~\eqref{qC2} and Eq.~\eqref{mgt} strongly depend on the selection rule of the transition operators, it is still essential to reevaluate the convergence properties of the individual quantities when studying the axial-vector process. The convergence of behaviors of $\delta_{C2}^A$ and $M_{GT}$ as a function of intermediate state number $N_\pi$ for the GT transition of $^{28}$P($\beta^+$)$^{28}$Si are illustrated in Fig.~\ref{conv}. In this case, $\delta_{C2}^A$ appears to converge at $N_\pi\approx 150-200$, whereas $M_{GT}$ still shows considerably variation until $N_\pi\approx 400$. Additionally, we have also observed that the rms radii calculated using the generalized expression\eqref{rad} converge much faster than $\delta_{C2}^A$.
\\

\begin{center}
\begin{figure}[ht!]
    \begin{tikzpicture}
     \begin{groupplot}[width=\linewidth, height=0.45\textwidth, group style={group size=1 by 1,xlabels at=edge bottom},
      ylabel=Conv. rates, xlabel=$N_{\pi}$, xmin=0,xmax=500, y unit=\%,grid=both]  
       \nextgroupplot[legend style={at={(0.95,0.95)}, anchor=north east,legend columns=1},legend cell align={left}]
       \addplot+[thick,error bars/.cd,y dir=both,y explicit] table[x=Npi,y=M] {convergence.dat}; 
       \addlegendentry{$R_{M_{GT}}$}
       \addplot+[thick,error bars/.cd,y dir=both,y explicit] table[x=Npi,y=C2] {convergence.dat}; 
       \addlegendentry{$R_{\delta_{C2}^A}$}
      \end{groupplot}
    \end{tikzpicture}
\caption{Convergence rates of $\delta_{C2}^A$ and $M_{GT}$ for the Gamow-Teller transition $^{28}$P$(\beta^+)^{28}$Si ($3_1^+ \to 2_1^+$) as functions of the intermediate state number $N_\pi$.}
\label{conv}
\end{figure}
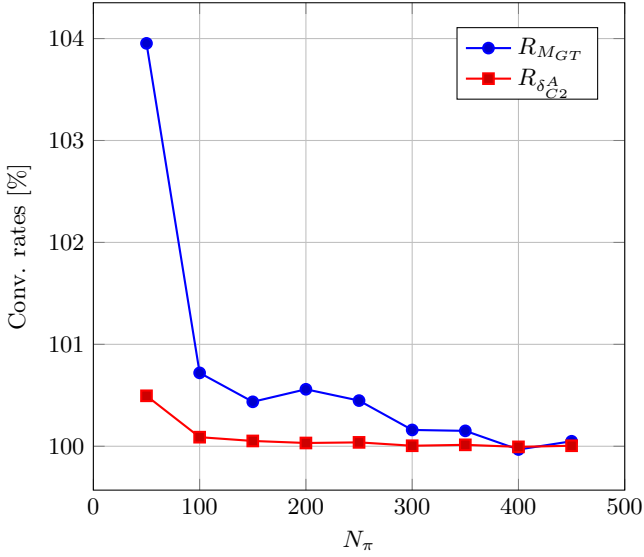
\end{center} 

\section{$\delta_{C2}$ in the isospin formalism}\label{iso}

The expression for $\delta_{C2}^A$, which was discussed in the preceding section, has been derived within the proton-neutron formalism. It is important to note that this correction term accounts only for the radial mismatch effect. Apart from $\Lambda_{ab}^\pi$, the other quantities in Eq.~\eqref{qC2} rely on the many-particle wave functions with definite isospin. Therefore, Eq.~\eqref{qC2} can be decomposed according to the possible isospin channels of intermediate states. This technique was employed to gain valuable insights into the structural behavior of $\delta_{C2}^V$ for the superallowed $0^+\rightarrow 0^+$ Fermi $\beta$ decay of isotriplets~\cite{ToHa2008}. An important observation from this study is the substantial cancellation between the isospin lesser and greater components of $\delta_{C2}^V$ when the corresponding radial mismatch factors do not differ significantly. Conversely, a notable difference in the radial mismatch factor can lead to a strong amplification of contributions from deeply bound orbits. For example, the inclusion of $sd$ orbits resulted in a significant improvement in the results for $^{46}$V($\beta^+$)$^{46}$Ti~\cite{ToHa2008}. 
To apply this technique to GT transitions, we begin by converting the proton and neutron spectroscopic amplitudes appearing in Eq.~\eqref{qC2} into isospin-invariant spectroscopic factors: 
\begin{equation}\label{spec}
\displaystyle A^T(x;\pi y) = C_y^{\pi x}\sqrt{S^T(x;\pi k_y)},
\end{equation} 
where $x=i$ or $f$ and $y=a$ or $b$. The spectroscopic factor $S^T(x;\pi k_y)$ does not depend on the isospin projection quantum numbers $\tau_y$. It is defined as: 
\begin{equation}
S^T(x;\pi k_y) =\frac{ (x||| a_{k_y}^\dagger |||\pi)^2 }{ (2J_x+1)(2T_x+1) },
\end{equation}
where the triple bar matrix element in this expression indicates that it is reduced in both coordinate space and isospin space.
Once again, we use round brackets for many-particle states with definite isospin. 

In the case of Fermi transitions, where the initial and final many-particle states are isobaric analogue states, we observe that $S^T(f;\pi k_a)=S^T(i;\pi k_b)$ with $k_a=k_b$. However, for GT transitions, this property does not generally hold due to the relaxation of isospin and angular momentum selection rules. An exception to this occurs in the GT transition between the isodoublet states of mirror nuclei in a single-$j$ configuration.

The sign of the spectroscopic amplitudes is determined solely by the isospin Clebsch-Gordan coefficient, given by $C_y^{\pi x}=\braket{T_\pi T_{\pi z}\frac{1}{2}\tau_{y}|T_xT_{xz}}$. This coefficient can be calculated using the following formula: 
\begin{equation}
\displaystyle C_y^{\pi x}=\pm\sqrt{ \frac{1}{2}\left(1\pm 2\tau_{y}\frac{T_{xz}}{T_\pi+\frac{1}{2}} \right) },
\end{equation}
where $T_\pi$ represents the isospin quantum number of the intermediate state. The $\pm$ sign corresponds to $T_\pi=T_x\mp\frac{1}{2}$, and the $\pm$ sign in front of the square root must be omitted if $\tau_{y}=-\frac{1}{2}$. It is important to note that we use $\tau$ for the isospin projection quantum number of the nucleon, deviating from the usual notation of $t_z$. 

Upon substituting Eq.\eqref{spec} into Eq.\eqref{qC2}, the general expression of $\delta_{C2}^A$ is reduced to:
\begin{equation}\label{CC2}
\begin{array}{ll}
\delta_{C2}^A &= \displaystyle \frac{2}{\mathcal{M}_{GT}} \sum_{k_ak_b\pi} \theta_{ab}^\lambda \Theta_{abfi}^{\pi\lambda} \Lambda_{ab}^\pi \xi_{ab} C_a^{\pi f} C_b^{\pi i} \\[0.15in]
&\times \sqrt{S^T(f;\pi k_a)S^T(i;\pi k_b)}, 
\end{array}
\end{equation}
this provides a more concise representation of $\delta_{C2}^A$ in terms of isospin-invariant spectroscopic factors. Note that the product of $\theta_{ab}^\lambda \Theta_{abfi}^{\pi\lambda}$ may have an inconsistent sign between the valence orbitals and the intermediate states, potentially resulting in a cancellation in $\delta_{C2}^A$. Further interesting properties of this correction term can be studied by examining specific isospin changes between the initial and final states. More details on this are discussed separately in the following subsections. 

\subsection{Gamow-Teller transition with $T_i=T_f$}

In the case of a GT transition with identical isospin quantum numbers for the initial and final many-particle states, the product of the isospin Clebsch-Gordan coefficients in Eq.~\eqref{CC2} can be explicitly written as: 
\begin{equation}\label{cg} 
\displaystyle C_a^{\pi f}C_b^{\pi i} = \pm \frac{1}{2\tilde{T}_\pi} \sqrt{ (\tilde{T}\pi \pm 2\tau_aT{iz}\pm 1)(\tilde{T}\pi \mp 2\tau_aT{iz}) } ,
\end{equation}
where $\tilde{T}\pi=T\pi+\frac{1}{2}$. It is important to note that all the $\pm$ signs in Eq.~\eqref{cg} (including the one in front of the square root) correspond to $T_\pi = T_i \pm \frac{1}{2}$. 
Substituting $T_\pi=T_i\pm\frac{1}{2}$ into Eq.~\eqref{cg}, then we obtain
\begin{equation}
C_a^{\pi f}C_b^{\pi i} = \frac{C_0}{2T_i}\left\{
\begin{array}{lll}
\displaystyle 1, & \text{if} & T_\pi = T_i - \frac{1}{2} \\[0.1in]
\displaystyle-\frac{T_i}{T_i+1}, & \text{if} & T_\pi = T_i + \frac{1}{2}, \\[0.1in]
\end{array}
\right.
\end{equation}
where $C_0=\sqrt{(T_i + 2\tau_aT_{iz} + 1)(T_i - 2\tau_aT_{iz})}$. Note that for $T_i=|T_{iz}|$, which holds true for most cases considered in the present study, we obtain $C_0=\sqrt{2T_i}$. Furthermore, it can be shown that $C_0$ is independent of the transition direction; thus, it does not affect the mirror asymmetry. 

Subsequently, the $\delta_{C2}^A$ expression can be separated into two parts, namely
\begin{widetext}
\begin{equation}\label{xxx}
\displaystyle 
\delta_{C2}^A = \frac{C_0}{T_i\mathcal{M}_{GT}} \sum_{k_ak_b}\theta_{ab}^\lambda\xi_{ab} \left[ \sum_{\pi'} \Theta_{abfi}^{\pi'\lambda} \Lambda_{ab}^{\pi'}\sqrt{S^T(f;\pi' k_a)S^T(i;\pi' k_b)} - \displaystyle\frac{T_i}{(T_i+1)} \sum_{\pi''} \Theta_{abfi}^{\pi''\lambda} \Lambda_{ab}^{\pi''}\sqrt{S^T(f;\pi'' k_a)S^T(i;\pi'' k_b)} \right], 
\end{equation}
\end{widetext}
where $\pi'$ and $\pi''$ denote an intermediate state with $T_\pi=T_i-\frac{1}{2}$ and $T_\pi=T_i+\frac{1}{2}$, respectively. 

The functions $\theta_{ab}^\lambda$ and $\Theta_{abfi}^{\pi\lambda}$ do not depend on the isospin of intermediate states. On average, the spectroscopic factors decrease with the excitation energy of the $(A-1)$-nucleon nucleus. Therefore, we can effectively assume that $S^T(f;\pi' k_a)S^T(i;\pi' k_b)>S^T(f;\pi'' k_a)S^T(i;\pi'' k_b)$ because the state $\ket{\pi''}$ lies at a higher energy compared to its isospin-lesser partner $\ket{\pi'}$. Similarly, a similar property can be expected for the radial mismatch factors, i.e., $\Lambda_{ab}^{\pi'}>\Lambda_{ab}^{\pi''}$ because the sensitivity of radial wave functions to the Coulomb and isovector potentials is diluted by increasing excitation energy of the intermediate nucleus. As a result, for $T_i=T_f$, there is generally a cancellation between the isospin-lesser and isospin-greater components of the radial mismatch correction term, as can be seen from Eq.\,\eqref{xxx}, 
particularly for deep-lying orbitals because of higher occupation probability. This cancellation behavior also appears in the case of superallowed $0^+\rightarrow 0^+$ Fermi $\beta$ decay of isotriplets\cite{ToHa2008,XaNa2022}. On the other hand, if the cancellation in spectroscopic factors is not so complete, the rest can be amplified by the radial mismatch factors which are sensitive to separation energies, enhancing $\delta_{C2}^A$. This effect was found to be important in $^{46}$V($\beta^+$)$^{46}$Ti~\cite{ToHa2008}. Therefore, experimental data on the spectroscopic factors would be an efficient indicator of core polarization, even though not precise enough to be used as input for $\delta_{C2}^A$ calculation. 

\subsection{Gamow-Teller transition with $T_i\ne T_f$}

For a GT transition with different isospin quantum numbers for the initial and final many-particle states, only intermediate states with $T_\pi = T+\frac{1}{2}$ (denoted as $\tilde{\pi}$) are allowed, where $T=\min{(T_i,T_f)}$. This restriction arises due to the isospin selection rule of GT transitions. In this case, the product of the isospin Clebsch-Gordan coefficients is greatly simplified as follows: 
\begin{equation}
\displaystyle C_a^{\tilde{\pi} f}C_b^{\tilde{\pi} i} = -\frac{\tilde{C}_0}{2(T+1)},
\end{equation}
where $\tilde{C}_0=\sqrt{ (T-2\tau_aT{iz})(T+2\tau_bT_{iz}+1) }$. Similar to $C_0$ in the previous subsection, $\tilde{C}_0$ does not depend on the transition direction. Furthermore, for $T_i=|T_{iz}|$ and $T_f=|T_{fz}|$, we obtain $\tilde{C}_0=(2T+1)$. 
Therefore, the expression of $\delta_{C2}^A$ is evaluated as 
\begin{equation}\label{equal}
\begin{array}{ll}
\delta_{C2}^A &= \displaystyle -\frac{\tilde{C}_0}{(T+1)\mathcal{M}_{GT}} \sum_{k_ak_b\tilde{\pi}}\theta_{ab}^\lambda \Theta_{abfi}^{\tilde{\pi}\lambda} \Lambda_{ab}^{\tilde{\pi}}\xi_{ab} \\[0.18in] 
& \displaystyle \times \sqrt{S^T(f;\tilde{\pi} k_a)S^T(i;\tilde{\pi} k_b)}.  
\end{array}
\end{equation}
We retain the $-$ sign in Eq.\,\eqref{equal}, which should be consistent and canceled out with the sign of $\mathcal{M}_{GT}$. Interestingly, Eq.\,\eqref{equal} consists of only a single isospin component, making $\delta_{C2}^A$ for GT transitions with $T_i\ne T_f$ generally more sensitive to core orbital contribution compared to other isospin configurations. Thus, if only the ISB effect is considered, the value of $\delta_{C2}^A$ for GT transitions with $T_i\ne T_f$ would be larger than that for GT transitions with $T_i = T_f$, particularly when $T_{\tilde{\pi}}=|T_{\pi z}|$, including the ground state, which yields a larger radial mismatch factor compared to greater values of $T_{\tilde{\pi}}$. Due to this property, shell model calculations of $\delta_{C2}^A$ for $T_i\ne T_f$ GT transitions generally require a larger model space. We also note that Fermi transitions with $T_i\ne T_f$ are isospin forbidden. 
All of the considered GT transitions in this work have $T_i\ne T_f$. 
In principle, the core-orbital contribution could be compensated by using an effective GT operator. A detailed study on this phenomenon will be presented in our subsequent paper. 

Note that only the effective interactions listed in Table~\ref{tb:2lev} are used in our calculations of the radial mismatch correction. We considered two methods for fitting the separation energies relative to the intermediate states: one by varying the depth of the volume term, $V_0$, and the other is by varying the strength of the surface-peaked term, $V_s$. In the latter method, the process begins by adjusting $V_0$ to reproduce the separation energies at the ground state of the $(A-1)$-nucleon system, then $V_s$ is adjusted for excited states. Our adopted values for $\delta_{C2}^A$ are taken as the averages of those obtained from the calculations with and without the surface-peaked term. Our numerical results for $\delta_{C2}^A$ are given in Table~\ref{resC2}. 

\begin{center}
\begin{figure*}[ht!]
    \begin{tikzpicture}
     \begin{groupplot}[width=\linewidth, height=0.45\textwidth, group style={group size=1 by 1,xlabels at=edge bottom},
      ylabel=Mirror asymmetry $\delta$, y unit=\%, xlabel=$A$ of mirror pairs, grid=both, xmin=7, xmax=36]  
       \nextgroupplot[legend style={at={(0.95,0.95)}, anchor=north east,legend columns=1},legend cell align={left}]
        \addplot+[thick,error bars/.cd,y dir=both,y explicit] table[x=A,y=Our, y error=std] {delta_calc.dat}; 
        \addlegendentry{This work}
        \addplot+[thick,error bars/.cd,y dir=both,y explicit] table[x=A,y=Smi2003] {delta_calc1.dat}; 
        \addlegendentry{SmVo2003}
        \addplot+[black,thick,error bars/.cd,y dir=both,y explicit] table[x=A,y=exp, y error=stdexp] {delta_exp.dat}; 
        \addlegendentry{Exp}
      \end{groupplot}
    \end{tikzpicture}
\caption{\label{zzz}(Color online) Comparison of mirror asymmetry values. The shell model results from a previous study (denoted as SmVo2003) are extracted from Ref.~\cite{Smirnova2003}. The experimental data are taken from Ref.\cite{Smirnova2003,PhysRevC.80.044302,sym13122278}. The uncertainty of our result is evaluated considering the charge radius data used to constrain the length parameter and the utilization of different effective interactions.}
\end{figure*}
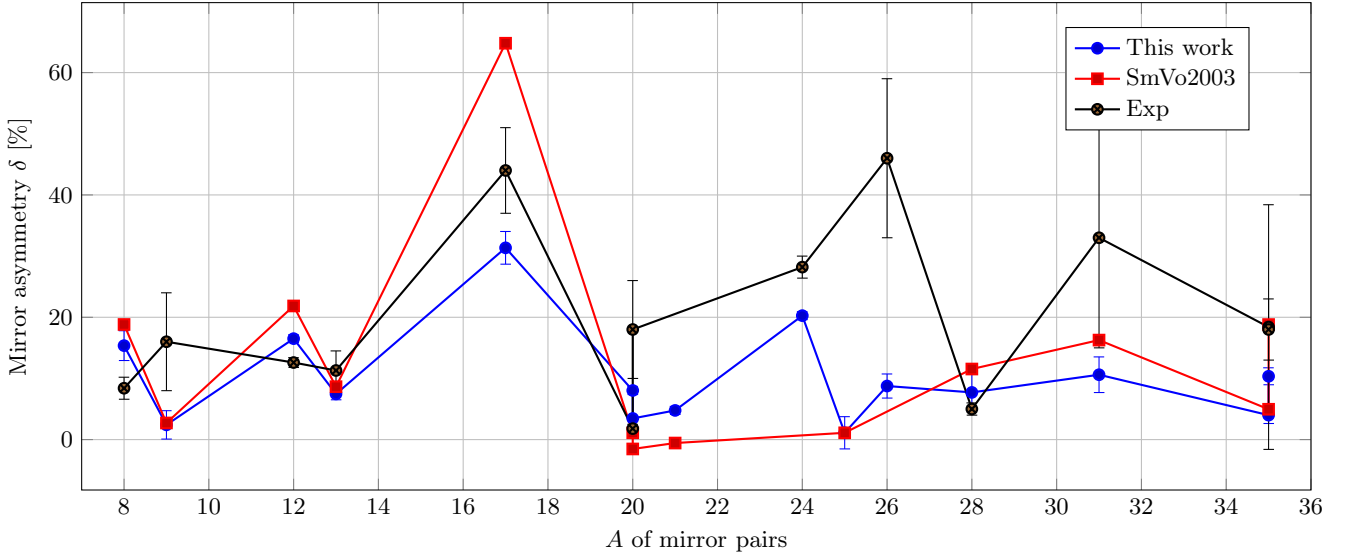
\end{center}


\section{Higher-order corrections}\label{x}

Recently, higher-order terms of $\delta_C^V$ were investigated for the superallowed $0^+\rightarrow 0^+$ Fermi $\beta$ decay of isotriplets through exact shell model calculations~\cite{xayavong2022higherorder}. The study revealed that the higher-order term contribution for this weak decay process is generally smaller than the theoretical uncertainty estimated from experimental data on charge radii used to constrain the length parameter of the WS potential and the use of different effective interactions. However, there is one exception, the heaviest emitter $^{74}$Rb, where the $2p_{1/2}$ orbital with a very low centrifugal barrier is close to the continuum. Given that the impact of isospin-symmetry breaking on GT transitions is expected to be stronger than on Fermi transitions, it becomes necessary to consider the higher-order terms for GT transitions. From Ref.~\cite{xayavong2022higherorder}, the next-to-leading order (NLO) term, $\delta_{C3}^A$ is generalized as follows, 
\begin{equation}\label{C3}
\displaystyle \delta_{C3}^A = -\delta_{C2} + \frac{2}{\mathcal{M}_{GT}} \sum_{k_ak_b\pi} \theta_{ab}^\lambda \Theta_{abfi}^{\pi\lambda} \Lambda_{ab}^{\pi}\xi_{ab} A(f;\pi a)A(i;\pi b), 
\end{equation} 
the structure of the second term in Eq.~\eqref{C3} appears very similar to that of $\delta_{C2}^A$, except that the isospin-invariant spectroscopic amplitudes are replaced with the isospin non-conserving ones (without superscript $T$). Therefore, by retaining $\delta_{C3}^A$, the contributions of isospin mixing and radial mismatch are no longer separable. Generally, the expression of $\delta_{C3}^A$ requires the same computational effort as that of $\delta_{C2}^A$; both of them necessitate a large number of intermediate states to achieve convergence. Fortunately, $\delta_{C3}$ has a smaller magnitude, allowing for a lower energy cut-off. 
\\
Another NLO term of $\delta_C^A$ is simply a function of the two LO terms. It is generalized as 
\begin{equation}\label{C4}
\delta_{C4}^A=\displaystyle -\frac{ \left(\delta_{C1}^A + \delta_{C2}^A \right)^2  }{4}. 
\end{equation}
Unlike $\delta_{C3}^A$, which can change its sign depending on the structure of a specific nucleus, Eq.\,\eqref{C4} indicates that the sign of $\delta_{C4}^A$ is always negative, leading to a destructive contribution to the total ISB correction. For the superallowed $0^+\rightarrow 0^+$ Fermi $\beta$ decay of isotriplets, the magnitude of $\delta_{C4}^V$ is generally smaller than that of $\delta_{C3}^V$~\cite{xayavong2022higherorder}. 
\\
Likewise, the next-to-next-to-leading order (NNLO) term can be evaluated with the successive order terms, namely 
\begin{equation}\label{NNLO} 
\delta_{C5}^A = \displaystyle -\frac{ ( \delta_{C1}^A + \delta_{C2}^A )\delta_{C3}^A }{2}. 
\end{equation} 
\\
The next-to-next-to-next-to-leading (NNNLO) order term is determined solely by $\delta_{C3}$ as follows: 
\begin{equation}
\delta_{C6}^A = \displaystyle -\frac{(\delta_{C3}^A)^2}{4}. 
\end{equation} 
Similar to $\delta_{C4}^A$, the NNNLO term is always negative, producing a destructive contribution to the total ISB correction. 

For simplicity, we use only one effective interaction for a given model space in our calculations of the higher-order terms, namely those listed in the fourth column of Table~\ref{tb:2lev}. Additionally, the surface-peaked term in the Woods-Saxon potential is not considered in the calculations of $\delta_{C3}^A$. 
Our numerical results for the higher-order terms are given in Table~\ref{higher}. It is observed that the higher-order terms are generally non-negligible for the GT transitions, except for $\delta_{C5}^A$ and $\delta_{C6}^A$. Remarkably, we found that $\delta_{C3}^A$ value for $^8$B($\beta^+$)$^8$Be is almost 3~\% and $\delta_{C4}^A$ value for $^{17}$Ne($\beta^+$)$^{17}$F is as large as approximately 7~\%. The corresponding contributions to the mirror asymmetry parameter of the GT transitions are given in Table~\ref{delta}.

\section{Mirror asymmetry}\label{c} 

The results obtained for the isospin-symmetry breaking correction terms have been utilized in the calculation of mirror asymmetry for the GT transition matrix elements. The computed asymmetry parameter values are presented in Table~\ref{delta}, alongside the earlier shell model calculations by Smirnova and Volpe~\cite{Smirnova2003} and experimental data~\cite{Smirnova2003,PhysRevC.80.044302,sym13122278}. The contributions of the LO and NLO correction terms are also separately provided in the same table. It is noticeable that the asymmetry $\delta$ displays significant fluctuations with respect to the mass number of the mirror pairs (see Fig.~\ref{zzz}). Our results show notable improvement in comparison to the previous shell model study and exhibit better agreement with experimental data. This suggests that the primary source of the mirror asymmetry $\delta$ lies in the isospin-symmetry breaking within nuclear structure. Recently, it was pointed out that the substantial mirror asymmetry 
($\delta=209\pm96$~\%) in GT transition $^{22}$Si($\beta^+$)$^{22}$Al and $^{22}$O($\beta^-$)$^{22}$F reveals the halo structure of $^{22}$Al~\cite{PhysRevLett.125.192503}. 
Similarly, the considerable mirror asymmetry of 44$\pm$7~\% observed between $^{17}$Ne($\beta^+$)$^{17}$F and $^{17}$N($\beta^-$)$^{17}$O reveals the halo structure of $^{17}$Ne. On the other hand, the significant asymmetry of 28.2$\pm$1.8~\% between $^{24}$Si($\beta^+$)$^{24}$Al and $^{24}$Ne($\beta^-$)$^{24}$Na was attribute to the changes in the proton $2s_{1/2}$ energy, also known as the Thomas-Ehrman shift~\cite{PhysRevC.80.044302}. This effect can also be interpreted as the influence of core polarization, as it is qualitatively consistent with our approach using the ZBM model space~\cite{PhysRevLett.21.39,PhysRevC.7.974}. It is likely that the $p$-orbital contribution is also significant for other nuclei in the lower to middle part of the $sd$ shell, as negative parity states often manifest at low energy in this mass region. Nonetheless, we discovered that the effective interactions from Ref.~\cite{ZBM1968,PhysRevC.7.974} proves inadequate in providing a reasonable description of energy levels for nuclei heavier than $A=24$. 
Additionally, while these interactions yield an exceptionally strong isospin mixing in both $^{22}$Si($\beta^+$)$^{22}$Al and $^{22}$O($\beta^-$)$^{22}$F transitions, as evident in Table~\ref{tb:2lev}, they fail to describe the large mirror asymmetry reported in Ref.~\cite{PhysRevLett.125.192503}. 
Hence, the development of new effective interactions tailored for the $p-sd$ cross-shell model space is imperative. 

However, considerable discrepancies between the calculated values and experimental data still persist. To decisively determine the presence of the second-class tensor current, it is imperative to address any shortcomings in the theoretical calculations. Potential sources of uncertainty in our theoretical approach include: 1) lack of experimental data for constraining radial wave functions, especially the data on charge and matter radii, 2) lack of accurate approach for the center-of-mass correction for the mean-field Woods-Saxon calculations, this correction could be critical in light nuclei like those with $A=8, 9, 12$, and $13$, and 3) the significance of core-orbital contributions for GT transitions with $T_i\ne T_f$. The last one requires a large model space or an effective operator instead of $\bm{\sigma}\cdot\bm{\tau}_\pm$. Furthermore, the contributions from radiative corrections $\delta_R'$ and $\delta_{NS}^A$, as well as second forbidden transitions, might also have a significant impact on the asymmetry $\delta$. 

Hence, the investigation of mirror asymmetry in GT transitions constitutes a substantial research endeavor. It serves as an interface bridging nuclear theory, responsible for determining the ISB corrections, nuclear experiments aimed at measuring the $ft$ values, and Quantum Electrodynamics (QED) which computes the radiative corrections. 

\section{Conclusion and Future Perspective}\label{con} 

In this comprehensive study, we have examined the influence of the isospin-symmetry breaking on Gamow-Teller transition matrix elements and their mirror asymmetry. Our approach involved the careful integration of a well-structured theoretical framework, which combines the shell model with isospin nonconserving effective interactions and Woods-Saxon radial wave functions, constrained by separation energies and charge radii when available. Through this framework, we have successfully dissected the isospin-symmetry breaking corrections into six distinct terms, allowing us to unveil its intricate components. At the leading order, these terms comprise two fundamental contributions: one that captures the isospin mixing within the many-particle shell model eigenfunctions, and another that accounts for the mismatch between the proton and neutron radial wave functions. Furthermore, our exploration extended to higher-order terms, which we accurately calculated. Through comprehensive calculations of the leading-order and higher-order terms, we have demonstrated their crucial role in refining the mirror asymmetry values for various GT transitions. Notably, our results exhibit a marked improvement over previous shell model studies, aligning more closely with experimental data. 

We have highlighted specific cases where mirror asymmetry serves as a powerful probe of nuclear structure. The large asymmetry, as observed between transitions in $^{17}$Ne and $^{17}$N, is sensitive to a unique nuclear phenomenon: the halo structure, similar to the case of $A=22$ as identified in Ref.~\cite{PhysRevLett.125.192503}. On the other hand, the pronounced asymmetry observed between $^{24}$Si($\beta^+$)$^{24}$Al and $^{24}$Ne($\beta^-$)$^{24}$Na transitions can be attributed to two distinct nuclear structure phenomena: one arising from the Thomas-Ehrman shift within the $sd$ valence space, and the other stemming from contributions of $p$-shell orbitals. 

However, despite the considerable strides our calculations have made, notable discrepancies between theoretical results and experimental data persist. Several aspects still need to be considered, including the necessity of experimental data on charge and matter radii to constrain the length parameters, the center-of-mass corrections in light nuclei, and the influence of radiative corrections and second forbidden transitions. Future research endeavors in this field will necessitate addressing these effects to refine our theoretical framework and enhance its predictive power. This research effort bridges the realms of nuclear theory, experimental measurements, Quantum Electrodynamics, and the exploration of weak interactions beyond the Standard Model.  

\begin{acknowledgments}
We thank N. A. Smirnova for fruitful discussion.
L. Xayavong and Y. Lim are supported by the National Research Foundation of Korea(NRF) grant funded by the Korea government(MSIT)(No. 2021R1A2C2094378). 
Y. Lim is also supported by the Yonsei University Research Fund of 2023-22-0126.
\end{acknowledgments}
%


\begin{thebibliography}{59}%
	\makeatletter
	\providecommand \@ifxundefined [1]{%
		\@ifx{#1\undefined}
	}%
	\providecommand \@ifnum [1]{%
		\ifnum #1\expandafter \@firstoftwo
		\else \expandafter \@secondoftwo
		\fi
	}%
	\providecommand \@ifx [1]{%
		\ifx #1\expandafter \@firstoftwo
		\else \expandafter \@secondoftwo
		\fi
	}%
	\providecommand \natexlab [1]{#1}%
	\providecommand \enquote  [1]{``#1''}%
	\providecommand \bibnamefont  [1]{#1}%
	\providecommand \bibfnamefont [1]{#1}%
	\providecommand \citenamefont [1]{#1}%
	\providecommand \href@noop [0]{\@secondoftwo}%
	\providecommand \href [0]{\begingroup \@sanitize@url \@href}%
	\providecommand \@href[1]{\@@startlink{#1}\@@href}%
	\providecommand \@@href[1]{\endgroup#1\@@endlink}%
	\providecommand \@sanitize@url [0]{\catcode `\\12\catcode `\$12\catcode
		`\&12\catcode `\#12\catcode `\^12\catcode `\_12\catcode `\%12\relax}%
	\providecommand \@@startlink[1]{}%
	\providecommand \@@endlink[0]{}%
	\providecommand \url  [0]{\begingroup\@sanitize@url \@url }%
	\providecommand \@url [1]{\endgroup\@href {#1}{\urlprefix }}%
	\providecommand \urlprefix  [0]{URL }%
	\providecommand \Eprint [0]{\href }%
	\providecommand \doibase [0]{https://doi.org/}%
	\providecommand \selectlanguage [0]{\@gobble}%
	\providecommand \bibinfo  [0]{\@secondoftwo}%
	\providecommand \bibfield  [0]{\@secondoftwo}%
	\providecommand \translation [1]{[#1]}%
	\providecommand \BibitemOpen [0]{}%
	\providecommand \bibitemStop [0]{}%
	\providecommand \bibitemNoStop [0]{.\EOS\space}%
	\providecommand \EOS [0]{\spacefactor3000\relax}%
	\providecommand \BibitemShut  [1]{\csname bibitem#1\endcsname}%
	\let\auto@bib@innerbib\@empty
	\bibitem [{\citenamefont {Hardy}\ and\ \citenamefont
		{Towner}(2020)}]{HaTo2020}%
	\BibitemOpen
	\bibfield  {author} {\bibinfo {author} {\bibfnamefont {J.~C.}\ \bibnamefont
			{Hardy}}\ and\ \bibinfo {author} {\bibfnamefont {I.~S.}\ \bibnamefont
			{Towner}},\ }\href {https://doi.org/10.1103/PhysRevC.102.045501} {\bibfield
		{journal} {\bibinfo  {journal} {Phys. Rev. C}\ }\textbf {\bibinfo {volume}
			{102}},\ \bibinfo {pages} {045501} (\bibinfo {year} {2020})}\BibitemShut
	{NoStop}%
	\bibitem [{\citenamefont {Webber}\ \emph {et~al.}(2011)\citenamefont {Webber}
		\emph {et~al.}}]{MuLan}%
	\BibitemOpen
	\bibfield  {author} {\bibinfo {author} {\bibfnamefont {D.~M.}\ \bibnamefont
			{Webber}} \emph {et~al.} (\bibinfo {collaboration} {MuLan Collaboration}),\
	}\href {https://doi.org/10.1103/PhysRevLett.106.041803} {\bibfield  {journal}
		{\bibinfo  {journal} {Phys. Rev. Lett.}\ }\textbf {\bibinfo {volume} {106}},\
		\bibinfo {pages} {041803} (\bibinfo {year} {2011})}\BibitemShut {NoStop}%
	\bibitem [{\citenamefont {Seng}(2021)}]{particles4040034}%
	\BibitemOpen
	\bibfield  {author} {\bibinfo {author} {\bibfnamefont {C.-Y.}\ \bibnamefont
			{Seng}},\ }\href {https://doi.org/10.3390/particles4040034} {\bibfield
		{journal} {\bibinfo  {journal} {Particles}\ }\textbf {\bibinfo {volume}
			{4}},\ \bibinfo {pages} {397} (\bibinfo {year} {2021})}\BibitemShut {NoStop}%
	\bibitem [{\citenamefont {Seng}\ and\ \citenamefont
		{Gorchtein}(2023)}]{PhysRevC.107.035503}%
	\BibitemOpen
	\bibfield  {author} {\bibinfo {author} {\bibfnamefont {C.-Y.}\ \bibnamefont
			{Seng}}\ and\ \bibinfo {author} {\bibfnamefont {M.}~\bibnamefont
			{Gorchtein}},\ }\href {https://doi.org/10.1103/PhysRevC.107.035503}
	{\bibfield  {journal} {\bibinfo  {journal} {Phys. Rev. C}\ }\textbf {\bibinfo
			{volume} {107}},\ \bibinfo {pages} {035503} (\bibinfo {year}
		{2023})}\BibitemShut {NoStop}%
	\bibitem [{\citenamefont {Towner}(1992)}]{TOWNER1992478}%
	\BibitemOpen
	\bibfield  {author} {\bibinfo {author} {\bibfnamefont {I.}~\bibnamefont
			{Towner}},\ }\href
	{https://doi.org/https://doi.org/10.1016/0375-9474(92)90170-O} {\bibfield
		{journal} {\bibinfo  {journal} {Nuclear Physics A}\ }\textbf {\bibinfo
			{volume} {540}},\ \bibinfo {pages} {478} (\bibinfo {year}
		{1992})}\BibitemShut {NoStop}%
	\bibitem [{\citenamefont {Towner}\ and\ \citenamefont
		{Hardy}(2008)}]{ToHa2008}%
	\BibitemOpen
	\bibfield  {author} {\bibinfo {author} {\bibfnamefont {I.~S.}\ \bibnamefont
			{Towner}}\ and\ \bibinfo {author} {\bibfnamefont {J.~C.}\ \bibnamefont
			{Hardy}},\ }\href {https://doi.org/10.1103/PhysRevC.77.025501} {\bibfield
		{journal} {\bibinfo  {journal} {Phys. Rev. C}\ }\textbf {\bibinfo {volume}
			{77}},\ \bibinfo {pages} {025501} (\bibinfo {year} {2008})}\BibitemShut
	{NoStop}%
	\bibitem [{\citenamefont {Towner}\ and\ \citenamefont
		{Hardy}(2010)}]{ToHa2010}%
	\BibitemOpen
	\bibfield  {author} {\bibinfo {author} {\bibfnamefont {I.~S.}\ \bibnamefont
			{Towner}}\ and\ \bibinfo {author} {\bibfnamefont {J.~C.}\ \bibnamefont
			{Hardy}},\ }\href {https://doi.org/10.1103/PhysRevC.82.065501} {\bibfield
		{journal} {\bibinfo  {journal} {Phys. Rev. C}\ }\textbf {\bibinfo {volume}
			{82}},\ \bibinfo {pages} {065501} (\bibinfo {year} {2010})}\BibitemShut
	{NoStop}%
	\bibitem [{\citenamefont {Smirnova}\ and\ \citenamefont
		{Volpe}(2003)}]{Smirnova2003}%
	\BibitemOpen
	\bibfield  {author} {\bibinfo {author} {\bibfnamefont {N.}~\bibnamefont
			{Smirnova}}\ and\ \bibinfo {author} {\bibfnamefont {C.}~\bibnamefont
			{Volpe}},\ }\href
	{https://doi.org/https://doi.org/10.1016/S0375-9474(02)01392-1} {\bibfield
		{journal} {\bibinfo  {journal} {Nuclear Physics A}\ }\textbf {\bibinfo
			{volume} {714}},\ \bibinfo {pages} {441} (\bibinfo {year}
		{2003})}\BibitemShut {NoStop}%
	\bibitem [{\citenamefont {Wrede}\ \emph {et~al.}(2010)\citenamefont {Wrede}
		\emph {et~al.}}]{PhysRevC.81.055503}%
	\BibitemOpen
	\bibfield  {author} {\bibinfo {author} {\bibfnamefont {C.}~\bibnamefont
			{Wrede}} \emph {et~al.},\ }\href {https://doi.org/10.1103/PhysRevC.81.055503}
	{\bibfield  {journal} {\bibinfo  {journal} {Phys. Rev. C}\ }\textbf {\bibinfo
			{volume} {81}},\ \bibinfo {pages} {055503} (\bibinfo {year}
		{2010})}\BibitemShut {NoStop}%
	\bibitem [{\citenamefont {Bhattacharya}\ \emph {et~al.}(2008)\citenamefont
		{Bhattacharya} \emph {et~al.}}]{Bhattacharya}%
	\BibitemOpen
	\bibfield  {author} {\bibinfo {author} {\bibfnamefont {M.}~\bibnamefont
			{Bhattacharya}} \emph {et~al.},\ }\href
	{https://doi.org/10.1103/PhysRevC.77.065503} {\bibfield  {journal} {\bibinfo
			{journal} {Phys. Rev. C}\ }\textbf {\bibinfo {volume} {77}},\ \bibinfo
		{pages} {065503} (\bibinfo {year} {2008})}\BibitemShut {NoStop}%
	\bibitem [{\citenamefont {Glassman}\ \emph {et~al.}(2019)\citenamefont
		{Glassman} \emph {et~al.}}]{glassman2019superallowed}%
	\BibitemOpen
	\bibfield  {author} {\bibinfo {author} {\bibfnamefont {B.~E.}\ \bibnamefont
			{Glassman}} \emph {et~al.},\ }\href@noop {} {\bibinfo {title} {Superallowed
			$0^+ \rightarrow 0^+$ $\beta$ decay of $t =2$ $^{20}$mg: $q_{\textrm{ec}}$
			value and $\beta\gamma$ branching}} (\bibinfo {year} {2019}),\ \Eprint
	{https://arxiv.org/abs/1910.12965} {arXiv:1910.12965 [nucl-ex]} \BibitemShut
	{NoStop}%
	\bibitem [{\citenamefont {Severijns}\ \emph {et~al.}(2023)\citenamefont
		{Severijns}, \citenamefont {Hayen}, \citenamefont {De~Leebeeck},
		\citenamefont {Vanlangendonck}, \citenamefont {Bodek}, \citenamefont
		{Rozpedzik},\ and\ \citenamefont {Towner}}]{PhysRevC.107.015502}%
	\BibitemOpen
	\bibfield  {author} {\bibinfo {author} {\bibfnamefont {N.}~\bibnamefont
			{Severijns}}, \bibinfo {author} {\bibfnamefont {L.}~\bibnamefont {Hayen}},
		\bibinfo {author} {\bibfnamefont {V.}~\bibnamefont {De~Leebeeck}}, \bibinfo
		{author} {\bibfnamefont {S.}~\bibnamefont {Vanlangendonck}}, \bibinfo
		{author} {\bibfnamefont {K.}~\bibnamefont {Bodek}}, \bibinfo {author}
		{\bibfnamefont {D.}~\bibnamefont {Rozpedzik}},\ and\ \bibinfo {author}
		{\bibfnamefont {I.~S.}\ \bibnamefont {Towner}},\ }\href
	{https://doi.org/10.1103/PhysRevC.107.015502} {\bibfield  {journal} {\bibinfo
			{journal} {Phys. Rev. C}\ }\textbf {\bibinfo {volume} {107}},\ \bibinfo
		{pages} {015502} (\bibinfo {year} {2023})}\BibitemShut {NoStop}%
	\bibitem [{\citenamefont {Gysbers}\ \emph {et~al.}(2019)\citenamefont
		{Gysbers}, \citenamefont {Hagen}, \citenamefont {Holt}, \citenamefont
		{Jansen}, \citenamefont {Morris}, \citenamefont {Navrátil}, \citenamefont
		{Papenbrock}, \citenamefont {Quaglioni}, \citenamefont {Schwenk},
		\citenamefont {Stroberg},\ and\ \citenamefont {Wendt}}]{GT}%
	\BibitemOpen
	\bibfield  {author} {\bibinfo {author} {\bibfnamefont {P.}~\bibnamefont
			{Gysbers}}, \bibinfo {author} {\bibfnamefont {G.}~\bibnamefont {Hagen}},
		\bibinfo {author} {\bibfnamefont {J.~D.}\ \bibnamefont {Holt}}, \bibinfo
		{author} {\bibfnamefont {G.~R.}\ \bibnamefont {Jansen}}, \bibinfo {author}
		{\bibfnamefont {T.~D.}\ \bibnamefont {Morris}}, \bibinfo {author}
		{\bibfnamefont {P.}~\bibnamefont {Navrátil}}, \bibinfo {author}
		{\bibfnamefont {T.}~\bibnamefont {Papenbrock}}, \bibinfo {author}
		{\bibfnamefont {S.}~\bibnamefont {Quaglioni}}, \bibinfo {author}
		{\bibfnamefont {A.}~\bibnamefont {Schwenk}}, \bibinfo {author} {\bibfnamefont
			{S.~R.}\ \bibnamefont {Stroberg}},\ and\ \bibinfo {author} {\bibfnamefont
			{K.~A.}\ \bibnamefont {Wendt}},\ }\href
	{https://doi.org/https://doi.org/10.1038/s41567-019-0450-7} {\bibfield
		{journal} {\bibinfo  {journal} {Nat. Phys.}\ }\textbf {\bibinfo {volume}
			{15}},\ \bibinfo {pages} {428} (\bibinfo {year} {2019})}\BibitemShut
	{NoStop}%
	\bibitem [{\citenamefont {Wilkinson}(1972)}]{WILKINSON1972289}%
	\BibitemOpen
	\bibfield  {author} {\bibinfo {author} {\bibfnamefont {D.}~\bibnamefont
			{Wilkinson}},\ }\href
	{https://doi.org/https://doi.org/10.1016/0375-9474(72)90370-3} {\bibfield
		{journal} {\bibinfo  {journal} {Nuclear Physics A}\ }\textbf {\bibinfo
			{volume} {179}},\ \bibinfo {pages} {289} (\bibinfo {year}
		{1972})}\BibitemShut {NoStop}%
	\bibitem [{\citenamefont {Barker}(1992)}]{BARKER1992147}%
	\BibitemOpen
	\bibfield  {author} {\bibinfo {author} {\bibfnamefont {F.}~\bibnamefont
			{Barker}},\ }\href
	{https://doi.org/https://doi.org/10.1016/0375-9474(92)90161-C} {\bibfield
		{journal} {\bibinfo  {journal} {Nuclear Physics A}\ }\textbf {\bibinfo
			{volume} {537}},\ \bibinfo {pages} {147} (\bibinfo {year}
		{1992})}\BibitemShut {NoStop}%
	\bibitem [{\citenamefont {Towner}(1973)}]{TOWNER1973589}%
	\BibitemOpen
	\bibfield  {author} {\bibinfo {author} {\bibfnamefont {I.}~\bibnamefont
			{Towner}},\ }\href
	{https://doi.org/https://doi.org/10.1016/0375-9474(73)90172-3} {\bibfield
		{journal} {\bibinfo  {journal} {Nuclear Physics A}\ }\textbf {\bibinfo
			{volume} {216}},\ \bibinfo {pages} {589} (\bibinfo {year}
		{1973})}\BibitemShut {NoStop}%
	\bibitem [{\citenamefont {Wilkinson}(1971)}]{PhysRevLett.27.1018}%
	\BibitemOpen
	\bibfield  {author} {\bibinfo {author} {\bibfnamefont {D.~H.}\ \bibnamefont
			{Wilkinson}},\ }\href {https://doi.org/10.1103/PhysRevLett.27.1018}
	{\bibfield  {journal} {\bibinfo  {journal} {Phys. Rev. Lett.}\ }\textbf
		{\bibinfo {volume} {27}},\ \bibinfo {pages} {1018} (\bibinfo {year}
		{1971})}\BibitemShut {NoStop}%
	\bibitem [{\citenamefont {Xayavong}\ and\ \citenamefont
		{Smirnova}(2022{\natexlab{a}})}]{xayavong2022higherorder}%
	\BibitemOpen
	\bibfield  {author} {\bibinfo {author} {\bibfnamefont {L.}~\bibnamefont
			{Xayavong}}\ and\ \bibinfo {author} {\bibfnamefont {N.}~\bibnamefont
			{Smirnova}},\ }\href@noop {} {\bibinfo {title} {Higher-order isospin-symmetry
			breaking corrections to nuclear matrix elements of superallowed $0^+\to 0^+$
			fermi $\beta$ decay of $t=1$ nuclei}} (\bibinfo {year}
	{2022}{\natexlab{a}}),\ \Eprint {https://arxiv.org/abs/2201.01035}
	{arXiv:2201.01035 [nucl-th]} \BibitemShut {NoStop}%
	\bibitem [{\citenamefont {Xayavong}\ and\ \citenamefont
		{Smirnova}(2018)}]{XaNa2018}%
	\BibitemOpen
	\bibfield  {author} {\bibinfo {author} {\bibfnamefont {L.}~\bibnamefont
			{Xayavong}}\ and\ \bibinfo {author} {\bibfnamefont {N.~A.}\ \bibnamefont
			{Smirnova}},\ }\href {https://doi.org/10.1103/PhysRevC.97.024324} {\bibfield
		{journal} {\bibinfo  {journal} {Phys. Rev. C}\ }\textbf {\bibinfo {volume}
			{97}},\ \bibinfo {pages} {024324} (\bibinfo {year} {2018})}\BibitemShut
	{NoStop}%
	\bibitem [{\citenamefont {M\"arkisch}\ \emph {et~al.}(2019)\citenamefont
		{M\"arkisch}, \citenamefont {Mest}, \citenamefont {Saul}, \citenamefont
		{Wang}, \citenamefont {Abele}, \citenamefont {Dubbers}, \citenamefont
		{Klopf}, \citenamefont {Petoukhov}, \citenamefont {Roick}, \citenamefont
		{Soldner},\ and\ \citenamefont {Werder}}]{PhysRevLett.122.242501}%
	\BibitemOpen
	\bibfield  {author} {\bibinfo {author} {\bibfnamefont {B.}~\bibnamefont
			{M\"arkisch}}, \bibinfo {author} {\bibfnamefont {H.}~\bibnamefont {Mest}},
		\bibinfo {author} {\bibfnamefont {H.}~\bibnamefont {Saul}}, \bibinfo {author}
		{\bibfnamefont {X.}~\bibnamefont {Wang}}, \bibinfo {author} {\bibfnamefont
			{H.}~\bibnamefont {Abele}}, \bibinfo {author} {\bibfnamefont
			{D.}~\bibnamefont {Dubbers}}, \bibinfo {author} {\bibfnamefont
			{M.}~\bibnamefont {Klopf}}, \bibinfo {author} {\bibfnamefont
			{A.}~\bibnamefont {Petoukhov}}, \bibinfo {author} {\bibfnamefont
			{C.}~\bibnamefont {Roick}}, \bibinfo {author} {\bibfnamefont
			{T.}~\bibnamefont {Soldner}},\ and\ \bibinfo {author} {\bibfnamefont
			{D.}~\bibnamefont {Werder}},\ }\href
	{https://doi.org/10.1103/PhysRevLett.122.242501} {\bibfield  {journal}
		{\bibinfo  {journal} {Phys. Rev. Lett.}\ }\textbf {\bibinfo {volume} {122}},\
		\bibinfo {pages} {242501} (\bibinfo {year} {2019})}\BibitemShut {NoStop}%
	\bibitem [{\citenamefont {Brown}\ and\ \citenamefont {Richter}(2006)}]{USDab}%
	\BibitemOpen
	\bibfield  {author} {\bibinfo {author} {\bibfnamefont {B.~A.}\ \bibnamefont
			{Brown}}\ and\ \bibinfo {author} {\bibfnamefont {W.~A.}\ \bibnamefont
			{Richter}},\ }\href {https://doi.org/10.1103/PhysRevC.74.034315} {\bibfield
		{journal} {\bibinfo  {journal} {Phys. Rev. C}\ }\textbf {\bibinfo {volume}
			{74}},\ \bibinfo {pages} {034315} (\bibinfo {year} {2006})}\BibitemShut
	{NoStop}%
	\bibitem [{\citenamefont {Xayavong}\ and\ \citenamefont
		{Smirnova}(2022{\natexlab{b}})}]{XaNa2022}%
	\BibitemOpen
	\bibfield  {author} {\bibinfo {author} {\bibfnamefont {L.}~\bibnamefont
			{Xayavong}}\ and\ \bibinfo {author} {\bibfnamefont {N.~A.}\ \bibnamefont
			{Smirnova}},\ }\href {https://doi.org/10.1103/PhysRevC.105.044308} {\bibfield
		{journal} {\bibinfo  {journal} {Phys. Rev. C}\ }\textbf {\bibinfo {volume}
			{105}},\ \bibinfo {pages} {044308} (\bibinfo {year}
		{2022}{\natexlab{b}})}\BibitemShut {NoStop}%
	\bibitem [{\citenamefont {Ormand}\ and\ \citenamefont
		{Brown}(1985)}]{OrBr1985}%
	\BibitemOpen
	\bibfield  {author} {\bibinfo {author} {\bibfnamefont {W.}~\bibnamefont
			{Ormand}}\ and\ \bibinfo {author} {\bibfnamefont {B.}~\bibnamefont {Brown}},\
	}\href {https://doi.org/https://doi.org/10.1016/0375-9474(85)90341-0}
	{\bibfield  {journal} {\bibinfo  {journal} {Nuclear Physics A}\ }\textbf
		{\bibinfo {volume} {440}},\ \bibinfo {pages} {274} (\bibinfo {year}
		{1985})}\BibitemShut {NoStop}%
	\bibitem [{\citenamefont {Severijns}\ \emph {et~al.}(2008)\citenamefont
		{Severijns}, \citenamefont {Tandecki}, \citenamefont {Phalet},\ and\
		\citenamefont {Towner}}]{Severijns2008}%
	\BibitemOpen
	\bibfield  {author} {\bibinfo {author} {\bibfnamefont {N.}~\bibnamefont
			{Severijns}}, \bibinfo {author} {\bibfnamefont {M.}~\bibnamefont {Tandecki}},
		\bibinfo {author} {\bibfnamefont {T.}~\bibnamefont {Phalet}},\ and\ \bibinfo
		{author} {\bibfnamefont {I.~S.}\ \bibnamefont {Towner}},\ }\href
	{https://doi.org/10.1103/PhysRevC.78.055501} {\bibfield  {journal} {\bibinfo
			{journal} {Phys. Rev. C}\ }\textbf {\bibinfo {volume} {78}},\ \bibinfo
		{pages} {055501} (\bibinfo {year} {2008})}\BibitemShut {NoStop}%
	\bibitem [{\citenamefont {Wildenthal}(1984)}]{USD}%
	\BibitemOpen
	\bibfield  {author} {\bibinfo {author} {\bibfnamefont {B.}~\bibnamefont
			{Wildenthal}},\ }\href
	{https://doi.org/https://doi.org/10.1016/0146-6410(84)90011-5} {\bibfield
		{journal} {\bibinfo  {journal} {Progress in Particle and Nuclear Physics}\
		}\textbf {\bibinfo {volume} {11}},\ \bibinfo {pages} {5} (\bibinfo {year}
		{1984})}\BibitemShut {NoStop}%
	\bibitem [{\citenamefont {Honma}\ \emph {et~al.}(2004)\citenamefont {Honma},
		\citenamefont {Otsuka}, \citenamefont {Brown},\ and\ \citenamefont
		{Mizusaki}}]{gx1a}%
	\BibitemOpen
	\bibfield  {author} {\bibinfo {author} {\bibfnamefont {M.}~\bibnamefont
			{Honma}}, \bibinfo {author} {\bibfnamefont {T.}~\bibnamefont {Otsuka}},
		\bibinfo {author} {\bibfnamefont {B.~A.}\ \bibnamefont {Brown}},\ and\
		\bibinfo {author} {\bibfnamefont {T.}~\bibnamefont {Mizusaki}},\ }\href
	{https://doi.org/10.1103/PhysRevC.69.034335} {\bibfield  {journal} {\bibinfo
			{journal} {Phys. Rev. C}\ }\textbf {\bibinfo {volume} {69}},\ \bibinfo
		{pages} {034335} (\bibinfo {year} {2004})}\BibitemShut {NoStop}%
	\bibitem [{\citenamefont {Poves}\ \emph {et~al.}(2001)\citenamefont {Poves},
		\citenamefont {Sánchez-Solano}, \citenamefont {Caurier},\ and\ \citenamefont
		{Nowacki}}]{kb3g}%
	\BibitemOpen
	\bibfield  {author} {\bibinfo {author} {\bibfnamefont {A.}~\bibnamefont
			{Poves}}, \bibinfo {author} {\bibfnamefont {J.}~\bibnamefont
			{Sánchez-Solano}}, \bibinfo {author} {\bibfnamefont {E.}~\bibnamefont
			{Caurier}},\ and\ \bibinfo {author} {\bibfnamefont {F.}~\bibnamefont
			{Nowacki}},\ }\href
	{https://doi.org/https://doi.org/10.1016/S0375-9474(01)00967-8} {\bibfield
		{journal} {\bibinfo  {journal} {Nuclear Physics A}\ }\textbf {\bibinfo
			{volume} {694}},\ \bibinfo {pages} {157} (\bibinfo {year}
		{2001})}\BibitemShut {NoStop}%
	\bibitem [{\citenamefont {Honma}\ \emph {et~al.}(2009)\citenamefont {Honma},
		\citenamefont {Otsuka}, \citenamefont {Mizusaki},\ and\ \citenamefont
		{Hjorth-Jensen}}]{jun45}%
	\BibitemOpen
	\bibfield  {author} {\bibinfo {author} {\bibfnamefont {M.}~\bibnamefont
			{Honma}}, \bibinfo {author} {\bibfnamefont {T.}~\bibnamefont {Otsuka}},
		\bibinfo {author} {\bibfnamefont {T.}~\bibnamefont {Mizusaki}},\ and\
		\bibinfo {author} {\bibfnamefont {M.}~\bibnamefont {Hjorth-Jensen}},\ }\href
	{https://doi.org/10.1103/PhysRevC.80.064323} {\bibfield  {journal} {\bibinfo
			{journal} {Phys. Rev. C}\ }\textbf {\bibinfo {volume} {80}},\ \bibinfo
		{pages} {064323} (\bibinfo {year} {2009})}\BibitemShut {NoStop}%
	\bibitem [{\citenamefont {Cohen}\ and\ \citenamefont
		{Kurath}(1965)}]{Cohen1965}%
	\BibitemOpen
	\bibfield  {author} {\bibinfo {author} {\bibfnamefont {S.}~\bibnamefont
			{Cohen}}\ and\ \bibinfo {author} {\bibfnamefont {D.}~\bibnamefont {Kurath}},\
	}\href {https://doi.org/https://doi.org/10.1016/0029-5582(65)90148-3}
	{\bibfield  {journal} {\bibinfo  {journal} {Nuclear Physics}\ }\textbf
		{\bibinfo {volume} {73}},\ \bibinfo {pages} {1} (\bibinfo {year}
		{1965})}\BibitemShut {NoStop}%
	\bibitem [{\citenamefont {Zuker}\ \emph
		{et~al.}(1968{\natexlab{a}})\citenamefont {Zuker}, \citenamefont {Buck},\
		and\ \citenamefont {McGrory}}]{ZBM1968}%
	\BibitemOpen
	\bibfield  {author} {\bibinfo {author} {\bibfnamefont {A.~P.}\ \bibnamefont
			{Zuker}}, \bibinfo {author} {\bibfnamefont {B.}~\bibnamefont {Buck}},\ and\
		\bibinfo {author} {\bibfnamefont {J.~B.}\ \bibnamefont {McGrory}},\ }\href
	{https://doi.org/10.1103/PhysRevLett.21.39} {\bibfield  {journal} {\bibinfo
			{journal} {Phys. Rev. Lett.}\ }\textbf {\bibinfo {volume} {21}},\ \bibinfo
		{pages} {39} (\bibinfo {year} {1968}{\natexlab{a}})}\BibitemShut {NoStop}%
	\bibitem [{\citenamefont {Ormand}\ and\ \citenamefont
		{Brown}(1989{\natexlab{a}})}]{OrBr1989}%
	\BibitemOpen
	\bibfield  {author} {\bibinfo {author} {\bibfnamefont {W.~E.}\ \bibnamefont
			{Ormand}}\ and\ \bibinfo {author} {\bibfnamefont {B.~A.}\ \bibnamefont
			{Brown}},\ }\href {https://doi.org/10.1103/PhysRevLett.62.866} {\bibfield
		{journal} {\bibinfo  {journal} {Phys. Rev. Lett.}\ }\textbf {\bibinfo
			{volume} {62}},\ \bibinfo {pages} {866} (\bibinfo {year}
		{1989}{\natexlab{a}})}\BibitemShut {NoStop}%
	\bibitem [{\citenamefont {Ormand}\ and\ \citenamefont
		{Brown}(1989{\natexlab{b}})}]{OrBr1989x}%
	\BibitemOpen
	\bibfield  {author} {\bibinfo {author} {\bibfnamefont {W.}~\bibnamefont
			{Ormand}}\ and\ \bibinfo {author} {\bibfnamefont {B.}~\bibnamefont {Brown}},\
	}\href {https://doi.org/https://doi.org/10.1016/0375-9474(89)90203-0}
	{\bibfield  {journal} {\bibinfo  {journal} {Nuclear Physics A}\ }\textbf
		{\bibinfo {volume} {491}},\ \bibinfo {pages} {1} (\bibinfo {year}
		{1989}{\natexlab{b}})}\BibitemShut {NoStop}%
	\bibitem [{\citenamefont {Ormand}\ and\ \citenamefont
		{Brown}(1995)}]{OrBr1995}%
	\BibitemOpen
	\bibfield  {author} {\bibinfo {author} {\bibfnamefont {W.~E.}\ \bibnamefont
			{Ormand}}\ and\ \bibinfo {author} {\bibfnamefont {B.~A.}\ \bibnamefont
			{Brown}},\ }\href {https://doi.org/10.1103/PhysRevC.52.2455} {\bibfield
		{journal} {\bibinfo  {journal} {Phys. Rev. C}\ }\textbf {\bibinfo {volume}
			{52}},\ \bibinfo {pages} {2455} (\bibinfo {year} {1995})}\BibitemShut
	{NoStop}%
	\bibitem [{\citenamefont {Lam}\ \emph {et~al.}(2013)\citenamefont {Lam},
		\citenamefont {Smirnova},\ and\ \citenamefont {Caurier}}]{Lam2013}%
	\BibitemOpen
	\bibfield  {author} {\bibinfo {author} {\bibfnamefont {Y.~H.}\ \bibnamefont
			{Lam}}, \bibinfo {author} {\bibfnamefont {N.~A.}\ \bibnamefont {Smirnova}},\
		and\ \bibinfo {author} {\bibfnamefont {E.}~\bibnamefont {Caurier}},\ }\href
	{https://doi.org/10.1103/PhysRevC.87.054304} {\bibfield  {journal} {\bibinfo
			{journal} {Phys. Rev. C}\ }\textbf {\bibinfo {volume} {87}},\ \bibinfo
		{pages} {054304} (\bibinfo {year} {2013})}\BibitemShut {NoStop}%
	\bibitem [{\citenamefont {Kirson}(2007)}]{Kir2006}%
	\BibitemOpen
	\bibfield  {author} {\bibinfo {author} {\bibfnamefont {M.~W.}\ \bibnamefont
			{Kirson}},\ }\href
	{https://doi.org/https://doi.org/10.1016/j.nuclphysa.2006.10.077} {\bibfield
		{journal} {\bibinfo  {journal} {Nuclear Physics A}\ }\textbf {\bibinfo
			{volume} {781}},\ \bibinfo {pages} {350} (\bibinfo {year}
		{2007})}\BibitemShut {NoStop}%
	\bibitem [{\citenamefont {Blomqvist}\ and\ \citenamefont
		{Molinari}(1968)}]{BLOMQVIST1968545}%
	\BibitemOpen
	\bibfield  {author} {\bibinfo {author} {\bibfnamefont {J.}~\bibnamefont
			{Blomqvist}}\ and\ \bibinfo {author} {\bibfnamefont {A.}~\bibnamefont
			{Molinari}},\ }\href
	{https://doi.org/https://doi.org/10.1016/0375-9474(68)90515-0} {\bibfield
		{journal} {\bibinfo  {journal} {Nuclear Physics A}\ }\textbf {\bibinfo
			{volume} {106}},\ \bibinfo {pages} {545} (\bibinfo {year}
		{1968})}\BibitemShut {NoStop}%
	\bibitem [{\citenamefont {Cohen}\ and\ \citenamefont
		{Kurath}(1967)}]{Cohen1967}%
	\BibitemOpen
	\bibfield  {author} {\bibinfo {author} {\bibfnamefont {S.}~\bibnamefont
			{Cohen}}\ and\ \bibinfo {author} {\bibfnamefont {D.}~\bibnamefont {Kurath}},\
	}\href {https://doi.org/https://doi.org/10.1016/0375-9474(67)90285-0}
	{\bibfield  {journal} {\bibinfo  {journal} {Nuclear Physics A}\ }\textbf
		{\bibinfo {volume} {101}},\ \bibinfo {pages} {1} (\bibinfo {year}
		{1967})}\BibitemShut {NoStop}%
	\bibitem [{\citenamefont {{Van Hees}}\ \emph {et~al.}(1988)\citenamefont {{Van
				Hees}}, \citenamefont {Wolters},\ and\ \citenamefont
		{Glaudemans}}]{VANHEES198861}%
	\BibitemOpen
	\bibfield  {author} {\bibinfo {author} {\bibfnamefont {A.}~\bibnamefont {{Van
					Hees}}}, \bibinfo {author} {\bibfnamefont {A.}~\bibnamefont {Wolters}},\ and\
		\bibinfo {author} {\bibfnamefont {P.}~\bibnamefont {Glaudemans}},\ }\href
	{https://doi.org/https://doi.org/10.1016/0375-9474(88)90373-9} {\bibfield
		{journal} {\bibinfo  {journal} {Nuclear Physics A}\ }\textbf {\bibinfo
			{volume} {476}},\ \bibinfo {pages} {61} (\bibinfo {year} {1988})}\BibitemShut
	{NoStop}%
	\bibitem [{\citenamefont {Julies}\ \emph {et~al.}(1992)\citenamefont {Julies},
		\citenamefont {Richter},\ and\ \citenamefont {Brown}}]{s.afr}%
	\BibitemOpen
	\bibfield  {author} {\bibinfo {author} {\bibfnamefont {R.~E.}\ \bibnamefont
			{Julies}}, \bibinfo {author} {\bibfnamefont {W.~A.}\ \bibnamefont
			{Richter}},\ and\ \bibinfo {author} {\bibfnamefont {B.~A.}\ \bibnamefont
			{Brown}},\ }\href@noop {} {\bibfield  {journal} {\bibinfo  {journal} {S. Afr.
				J. Phys.}\ }\textbf {\bibinfo {volume} {15}},\ \bibinfo {pages} {3/4}
		(\bibinfo {year} {1992})}\BibitemShut {NoStop}%
	\bibitem [{\citenamefont {McGrory}\ and\ \citenamefont
		{Wildenthal}(1973)}]{PhysRevC.7.974}%
	\BibitemOpen
	\bibfield  {author} {\bibinfo {author} {\bibfnamefont {J.~B.}\ \bibnamefont
			{McGrory}}\ and\ \bibinfo {author} {\bibfnamefont {B.~H.}\ \bibnamefont
			{Wildenthal}},\ }\href {https://doi.org/10.1103/PhysRevC.7.974} {\bibfield
		{journal} {\bibinfo  {journal} {Phys. Rev. C}\ }\textbf {\bibinfo {volume}
			{7}},\ \bibinfo {pages} {974} (\bibinfo {year} {1973})}\BibitemShut {NoStop}%
	\bibitem [{\citenamefont {Brown}\ and\ \citenamefont {Rae}(2014)}]{NuShellX}%
	\BibitemOpen
	\bibfield  {author} {\bibinfo {author} {\bibfnamefont {B.}~\bibnamefont
			{Brown}}\ and\ \bibinfo {author} {\bibfnamefont {W.}~\bibnamefont {Rae}},\
	}\href {https://doi.org/https://doi.org/10.1016/j.nds.2014.07.022} {\bibfield
		{journal} {\bibinfo  {journal} {Nuclear Data Sheets}\ }\textbf {\bibinfo
			{volume} {120}},\ \bibinfo {pages} {115} (\bibinfo {year}
		{2014})}\BibitemShut {NoStop}%
	\bibitem [{\citenamefont {Ichikawa}\ \emph {et~al.}(2009)\citenamefont
		{Ichikawa} \emph {et~al.}}]{PhysRevC.80.044302}%
	\BibitemOpen
	\bibfield  {author} {\bibinfo {author} {\bibfnamefont {Y.}~\bibnamefont
			{Ichikawa}} \emph {et~al.},\ }\href
	{https://doi.org/10.1103/PhysRevC.80.044302} {\bibfield  {journal} {\bibinfo
			{journal} {Phys. Rev. C}\ }\textbf {\bibinfo {volume} {80}},\ \bibinfo
		{pages} {044302} (\bibinfo {year} {2009})}\BibitemShut {NoStop}%
	\bibitem [{\citenamefont {Jian}\ \emph {et~al.}(2021)\citenamefont {Jian} \emph
		{et~al.}}]{sym13122278}%
	\BibitemOpen
	\bibfield  {author} {\bibinfo {author} {\bibfnamefont {H.}~\bibnamefont
			{Jian}} \emph {et~al.},\ }\bibfield  {journal} {\bibinfo  {journal}
		{Symmetry}\ }\textbf {\bibinfo {volume} {13}},\ \href
	{https://doi.org/10.3390/sym13122278} {10.3390/sym13122278} (\bibinfo {year}
	{2021})\BibitemShut {NoStop}%
	\bibitem [{\citenamefont {Thomas}(1952)}]{PhysRev.88.1109}%
	\BibitemOpen
	\bibfield  {author} {\bibinfo {author} {\bibfnamefont {R.~G.}\ \bibnamefont
			{Thomas}},\ }\href {https://doi.org/10.1103/PhysRev.88.1109} {\bibfield
		{journal} {\bibinfo  {journal} {Phys. Rev.}\ }\textbf {\bibinfo {volume}
			{88}},\ \bibinfo {pages} {1109} (\bibinfo {year} {1952})}\BibitemShut
	{NoStop}%
	\bibitem [{\citenamefont {Ehrman}(1951)}]{PhysRev.81.412}%
	\BibitemOpen
	\bibfield  {author} {\bibinfo {author} {\bibfnamefont {J.~B.}\ \bibnamefont
			{Ehrman}},\ }\href {https://doi.org/10.1103/PhysRev.81.412} {\bibfield
		{journal} {\bibinfo  {journal} {Phys. Rev.}\ }\textbf {\bibinfo {volume}
			{81}},\ \bibinfo {pages} {412} (\bibinfo {year} {1951})}\BibitemShut
	{NoStop}%
	\bibitem [{\citenamefont {Zuker}\ \emph
		{et~al.}(1968{\natexlab{b}})\citenamefont {Zuker}, \citenamefont {Buck},\
		and\ \citenamefont {McGrory}}]{PhysRevLett.21.39}%
	\BibitemOpen
	\bibfield  {author} {\bibinfo {author} {\bibfnamefont {A.~P.}\ \bibnamefont
			{Zuker}}, \bibinfo {author} {\bibfnamefont {B.}~\bibnamefont {Buck}},\ and\
		\bibinfo {author} {\bibfnamefont {J.~B.}\ \bibnamefont {McGrory}},\ }\href
	{https://doi.org/10.1103/PhysRevLett.21.39} {\bibfield  {journal} {\bibinfo
			{journal} {Phys. Rev. Lett.}\ }\textbf {\bibinfo {volume} {21}},\ \bibinfo
		{pages} {39} (\bibinfo {year} {1968}{\natexlab{b}})}\BibitemShut {NoStop}%
	\bibitem [{\citenamefont {Audi}\ \emph {et~al.}(2012)\citenamefont {Audi},
		\citenamefont {Kondev}, \citenamefont {Wang}, \citenamefont {Pfeiffer},
		\citenamefont {Sun}, \citenamefont {Blachot},\ and\ \citenamefont
		{MacCormick}}]{AME2012}%
	\BibitemOpen
	\bibfield  {author} {\bibinfo {author} {\bibfnamefont {G.}~\bibnamefont
			{Audi}}, \bibinfo {author} {\bibfnamefont {F.}~\bibnamefont {Kondev}},
		\bibinfo {author} {\bibfnamefont {M.}~\bibnamefont {Wang}}, \bibinfo {author}
		{\bibfnamefont {B.}~\bibnamefont {Pfeiffer}}, \bibinfo {author}
		{\bibfnamefont {X.}~\bibnamefont {Sun}}, \bibinfo {author} {\bibfnamefont
			{J.}~\bibnamefont {Blachot}},\ and\ \bibinfo {author} {\bibfnamefont
			{M.}~\bibnamefont {MacCormick}},\ }\href
	{https://doi.org/10.1088/1674-1137/36/12/001} {\bibfield  {journal} {\bibinfo
			{journal} {Chinese Physics C}\ }\textbf {\bibinfo {volume} {36}},\ \bibinfo
		{pages} {1157} (\bibinfo {year} {2012})}\BibitemShut {NoStop}%
	\bibitem [{\citenamefont {Bohr}\ and\ \citenamefont
		{Mottelson}(1998)}]{BohrMott}%
	\BibitemOpen
	\bibfield  {author} {\bibinfo {author} {\bibfnamefont {A.}~\bibnamefont
			{Bohr}}\ and\ \bibinfo {author} {\bibfnamefont {B.~R.}\ \bibnamefont
			{Mottelson}},\ }\href {https://doi.org/10.1142/3530} {\emph {\bibinfo {title}
			{Nuclear Structure}}}\ (\bibinfo  {publisher} {World Scientific Publishing
		Company},\ \bibinfo {year} {1998})\ \Eprint
	{https://arxiv.org/abs/https://www.worldscientific.com/doi/pdf/10.1142/3530}
	{https://www.worldscientific.com/doi/pdf/10.1142/3530} \BibitemShut {NoStop}%
	\bibitem [{\citenamefont {Schwierz}\ \emph {et~al.}(2007)\citenamefont
		{Schwierz}, \citenamefont {Wiedenhöver},\ and\ \citenamefont {Volya}}]{SWV}%
	\BibitemOpen
	\bibfield  {author} {\bibinfo {author} {\bibfnamefont {N.}~\bibnamefont
			{Schwierz}}, \bibinfo {author} {\bibfnamefont {I.}~\bibnamefont
			{Wiedenhöver}},\ and\ \bibinfo {author} {\bibfnamefont {A.}~\bibnamefont
			{Volya}},\ }\href@noop {} {\bibfield  {journal} {\bibinfo  {journal}
			{arXiv:0709.3525v1 [nucl-th]}\ } (\bibinfo {year} {2007})}\BibitemShut
	{NoStop}%
	\bibitem [{\citenamefont {Elton}(1961)}]{Elton}%
	\BibitemOpen
	\bibfield  {author} {\bibinfo {author} {\bibfnamefont {L.}~\bibnamefont
			{Elton}},\ }\href@noop {} {\bibfield  {journal} {\bibinfo  {journal} {Nuclear
				sizes (Oxford University Press)}\ } (\bibinfo {year} {1961})}\BibitemShut
	{NoStop}%
	\bibitem [{\citenamefont {Pinkston}\ and\ \citenamefont
		{Satchler}(1965)}]{PINKSTON1965641}%
	\BibitemOpen
	\bibfield  {author} {\bibinfo {author} {\bibfnamefont {W.}~\bibnamefont
			{Pinkston}}\ and\ \bibinfo {author} {\bibfnamefont {G.}~\bibnamefont
			{Satchler}},\ }\href
	{https://doi.org/https://doi.org/10.1016/0029-5582(65)90417-7} {\bibfield
		{journal} {\bibinfo  {journal} {Nuclear Physics}\ }\textbf {\bibinfo {volume}
			{72}},\ \bibinfo {pages} {641} (\bibinfo {year} {1965})}\BibitemShut
	{NoStop}%
	\bibitem [{\citenamefont {Mayer}(1950{\natexlab{a}})}]{PhysRev.78.16}%
	\BibitemOpen
	\bibfield  {author} {\bibinfo {author} {\bibfnamefont {M.~G.}\ \bibnamefont
			{Mayer}},\ }\href {https://doi.org/10.1103/PhysRev.78.16} {\bibfield
		{journal} {\bibinfo  {journal} {Phys. Rev.}\ }\textbf {\bibinfo {volume}
			{78}},\ \bibinfo {pages} {16} (\bibinfo {year}
		{1950}{\natexlab{a}})}\BibitemShut {NoStop}%
	\bibitem [{\citenamefont {Mayer}(1950{\natexlab{b}})}]{PhysRev.78.22}%
	\BibitemOpen
	\bibfield  {author} {\bibinfo {author} {\bibfnamefont {M.~G.}\ \bibnamefont
			{Mayer}},\ }\href {https://doi.org/10.1103/PhysRev.78.22} {\bibfield
		{journal} {\bibinfo  {journal} {Phys. Rev.}\ }\textbf {\bibinfo {volume}
			{78}},\ \bibinfo {pages} {22} (\bibinfo {year}
		{1950}{\natexlab{b}})}\BibitemShut {NoStop}%
	\bibitem [{\citenamefont {Thomas}(1926)}]{Thomas}%
	\BibitemOpen
	\bibfield  {author} {\bibinfo {author} {\bibfnamefont {L.}~\bibnamefont
			{Thomas}},\ }\bibfield  {journal} {\bibinfo  {journal} {Nature {\bf117},
			514}\ }\href {https://doi.org/10.1038/117514a0} {10.1038/117514a0} (\bibinfo
	{year} {1926})\BibitemShut {NoStop}%
	\bibitem [{\citenamefont {Horiuchi}(2021)}]{10.1093/ptep/ptab136}%
	\BibitemOpen
	\bibfield  {author} {\bibinfo {author} {\bibfnamefont {W.}~\bibnamefont
			{Horiuchi}},\ }\bibfield  {journal} {\bibinfo  {journal} {Progress of
			Theoretical and Experimental Physics}\ }\textbf {\bibinfo {volume} {2021}},\
	\href {https://doi.org/10.1093/ptep/ptab136} {10.1093/ptep/ptab136} (\bibinfo
	{year} {2021}),\ \bibinfo {note} {123D01},\ \Eprint
	{https://arxiv.org/abs/https://academic.oup.com/ptep/article-pdf/2021/12/123D01/42899146/ptab136.pdf}
	{https://academic.oup.com/ptep/article-pdf/2021/12/123D01/42899146/ptab136.pdf}
	\BibitemShut {NoStop}%
	\bibitem [{\citenamefont {Xayavong}(2016)}]{Xthesis}%
	\BibitemOpen
	\bibfield  {author} {\bibinfo {author} {\bibfnamefont {L.}~\bibnamefont
			{Xayavong}},\ }\emph {\bibinfo {title} {{Calculs th{\'e}oriques de
				corrections nucl{\'e}aires aux taux de transitions $\beta$ super-permises
				pour les tests du Mod{\`e}le Standard}}},\ \href
	{https://theses.hal.science/tel-01674248} {\bibinfo {type} {Theses}},\
	\bibinfo  {school} {{Universit{\'e} de Bordeaux}} (\bibinfo {year}
	{2016})\BibitemShut {NoStop}%
	\bibitem [{\citenamefont {Wimmer}\ \emph {et~al.}(2021)\citenamefont {Wimmer}
		\emph {et~al.}}]{PhysRevLett.126.072501}%
	\BibitemOpen
	\bibfield  {author} {\bibinfo {author} {\bibfnamefont {K.}~\bibnamefont
			{Wimmer}} \emph {et~al.},\ }\href
	{https://doi.org/10.1103/PhysRevLett.126.072501} {\bibfield  {journal}
		{\bibinfo  {journal} {Phys. Rev. Lett.}\ }\textbf {\bibinfo {volume} {126}},\
		\bibinfo {pages} {072501} (\bibinfo {year} {2021})}\BibitemShut {NoStop}%
	\bibitem [{\citenamefont {Towner}\ and\ \citenamefont
		{Hardy}(2002)}]{ToHa2002}%
	\BibitemOpen
	\bibfield  {author} {\bibinfo {author} {\bibfnamefont {I.~S.}\ \bibnamefont
			{Towner}}\ and\ \bibinfo {author} {\bibfnamefont {J.~C.}\ \bibnamefont
			{Hardy}},\ }\href {https://doi.org/10.1103/PhysRevC.66.035501} {\bibfield
		{journal} {\bibinfo  {journal} {Phys. Rev. C}\ }\textbf {\bibinfo {volume}
			{66}},\ \bibinfo {pages} {035501} (\bibinfo {year} {2002})}\BibitemShut
	{NoStop}%
	\bibitem [{\citenamefont {Lee}\ \emph {et~al.}(2020)\citenamefont {Lee} \emph
		{et~al.}}]{PhysRevLett.125.192503}%
	\BibitemOpen
	\bibfield  {author} {\bibinfo {author} {\bibfnamefont {J.}~\bibnamefont
			{Lee}} \emph {et~al.} (\bibinfo {collaboration} {RIBLL Collaboration}),\
	}\href {https://doi.org/10.1103/PhysRevLett.125.192503} {\bibfield  {journal}
		{\bibinfo  {journal} {Phys. Rev. Lett.}\ }\textbf {\bibinfo {volume} {125}},\
		\bibinfo {pages} {192503} (\bibinfo {year} {2020})}\BibitemShut {NoStop}%
\end{thebibliography}
%

\appendix*
\section{Numerical results} 

\begin{table*}[ht!]
\caption{Numerical results for the isospin-mixing correction using different effective interactions are presented. The isospin-symmetry Gamow-Teller matrix elements for the first and second admixed states are also listed. The + (-) sign corresponds to the $\beta^+$ ($\beta^-$) decay of a given pair of Gamow-Teller transitions. The units of $\delta_{C1}^-$, $\delta_{C1}^+$, and $\delta_1^{ISB}$ are in percentage (\%).}
\begin{ruledtabular}
\label{resC1}
\begin{tabular}{c|c|c|c|c|c|c|c|c}
Mirror Pair	&	$J_i^\pi	T_i$	&	$J_f^\pi	T_f$	&	Interaction	&	$\mathcal{M}_0$	&	$\mathcal{M}_1$	&	$\delta_{C1}^-$	&	$\delta_{C1}^+$	&	$\delta_1^{ISB}$	\\
\hline			
$^8$Li$(\beta^-)^8$Be; $^8$B($\beta^+$)$^8$Be	&	$2^+	1$	&	$2_1^+	0$	&	CKPOT/CD	&	0.580	&	3.599	&	-10.775	&	-7.597	&	3.178	\\
	&			&			&	CKI/CD	&	0.333	&	3.433	&	-11.070	&	-9.059	&	2.011	\\
	&			&			&	CKII/CD	&	0.425	&	3.598	&	-12.643	&	-9.673	&	2.970	\\
	&			&			&	PJP/CD	&	0.448	&	3.541	&	-9.176	&	-7.014	&	2.162	\\
	&			&			&	PJT/CD	&	0.413	&	3.447	&	-9.037	&	-6.768	&	2.269	\\
	&			&			&	PMOM/CD	&	0.677	&	3.381	&	-7.148	&	-4.844	&	2.304	\\
	&			&			&	Average :	&	0.479$\pm$0.115	&	3.500$\pm$0.084	&	-9.975$\pm$1.754	&	-7.493$\pm$1.580	&	2.482$\pm$0.433	\\
\hline				
$^9$Li($\beta^-$)$^9$Be; $^9$C($\beta^+$)$^9$B	&	$\frac{3}{2}^-	\frac{3}{2}$	&	$\frac{3}{2}_1^-	\frac{3}{2}$	&	CKPOT/CD	&	0.60438	&	0.58574	&	-4.785	&	-6.082	&	-1.297	\\
	&			&			&	CKI/CD	&	0.38464	&	0.14028	&	-5.948	&	-6.349	&	-0.400	\\
	&			&			&	CKII/CD	&	0.56331	&	0.31466	&	-4.772	&	-5.272	&	-0.501	\\
	&			&			&	PJP/CD	&	0.37931	&	0.46563	&	-5.510	&	-7.709	&	-2.199	\\
	&			&			&	PJT/CD	&	0.42677	&	0.19884	&	-6.163	&	-5.539	&	0.623	\\
	&			&			&	PMOM/CD	&	0.64342	&	0.4727	&	-3.603	&	-3.320	&	0.283	\\
	&			&			&	Average :	&	0.500$\pm$0.117	&	0.363$\pm$0.174	&	-5.130$\pm$0.945	&	-5.712$\pm$1.447	&	-0.582$\pm$1.036	\\
\hline				
$^{12}$B($\beta^-$)$^{12}$C; $^{12}$N($\beta^+$)$^{12}$C	&	$1^+	1$	&	$0_1^+	0$	&	CKPOT/CD	&	0.960	&	1.893	&	-5.602	&	-4.926	&	0.675	\\
	&			&			&	CKI/CD	&	0.964	&	1.915	&	-5.662	&	-4.415	&	1.247	\\
	&			&			&	CKII/CD	&	0.996	&	1.865	&	-6.033	&	-5.014	&	1.018	\\
	&			&			&	PJP/CD	&	0.920	&	1.920	&	-4.840	&	-4.147	&	0.694	\\
	&			&			&	PJT/CD	&	0.982	&	1.802	&	-5.033	&	-4.098	&	0.935	\\
	&			&			&	PMOM/CD	&	1.064	&	1.837	&	-4.300	&	-3.452	&	0.848	\\
	&			&			&	Average :	&	0.981$\pm$0.029	&	1.872$\pm$0.046	&	-5.245$\pm$0.636	&	-4.342$\pm$0.581	&	0.903$\pm$0.215	\\
\hline					
$^{13}$B($\beta^-$)$^{13}$C; $^{13}$O($\beta^+$)$^{13}$N	&	$\frac{3}{2}^-	\frac{3}{2}$	&	$\frac{1}{2}_1^-	\frac{1}{2}$	&	CKPOT/CD	&	1.376	&	-0.462	&	-3.589	&	-1.489	&	2.101	\\
	&			&			&	CKI/CD	&	1.447	&	0.811	&	-3.752	&	-1.086	&	2.665	\\
	&			&			&	CKII/CD	&	1.428	&	0.796	&	-3.795	&	-1.203	&	2.592	\\
	&			&			&	PJP/CD	&	1.316	&	0.422	&	-3.838	&	-1.628	&	2.210	\\
	&			&			&	PJT/CD	&	1.356	&	0.725	&	-4.105	&	-1.373	&	2.732	\\
	&			&			&	PMOM/CD	&	1.421	&	0.580	&	-3.230	&	-1.123	&	2.107	\\
	&			&			&	Average :	&	1.391$\pm$0.050	&	-0.227$\pm$0.669	&	-3.718$\pm$0.292	&	-1.317$\pm$0.216	&	2.401$\pm$0.293	\\
\hline				
$^{17}$N($\beta^-$)$^{17}$O; $^{17}$Ne($\beta^+$)$^{17}$F	&	$\frac{1}{2}^-	\frac{3}{2}$	&	$\frac{3}{2}_1^-	\frac{1}{2}$	&	REWIL/CD	&	0.553	&	1.952	&	13.602	&	49.436	&	35.834	\\
\hline					
$^{20}$F($\beta^-$)$^{20}$Ne; $^{20}$Na($\beta^+$)$^{20}$Ne	&	$2^+	1$	&	$2_1^+	0$	&	REWIL/CD	&	0.882	&	1.029	&	6.156	&	12.516	&	6.36	\\
\hline					
$^{20}$O($\beta^-$)$^{20}$F; $^{20}$Mg($\beta^+$)$^{20}$Na	&	$0^+	2$	&	$1_1^+	1$	&	REWIL/CD	&	1.37	&	1.017	&	9.079	&	11.32	&	2.241	\\
\hline					
$^{21}$F($\beta^-$)$^{21}$Ne; $^{21}$Mg($\beta^+$)$^{21}$Na	&	$\frac{5}{2}^+	\frac{3}{2}$	&	$\frac{3}{2}_1^+	\frac{1}{2}$	&	REWIL/CD	&	0.452	&	1.058	&	5.967	&	8.775	&	2.808	\\
\hline
$^{22}$O($\beta^-$)$^{22}$F; $^{22}$Si($\beta^+$)$^{22}$Al	&	$0^+	3$	&	$1_1^+	2$	&	REWIL/CD	&	0.245	&	1.878	&	164.147	&	127.780	&	41.906	\\
\hline	
$^{24}$Ne($\beta^-$)$^{24}$Na; $^{24}$Si($\beta^+$)$^{24}$Al	&	$0^+	2$	&	$1_1^+	1$	&	REWIL/CD	&	0.877	&	1.513	&	0.694	&	19.496	&	18.802	\\
\hline				
$^{25}$Na($\beta^-$)$^{25}$Mg; $^{25}$Si($\beta^+$)$^{25}$Al	&	$\frac{5}{2}^+	\frac{3}{2}$	&	$\frac{5}{2}_1^+	\frac{1}{2}$	&	USDA/CD	&	0.406	&	0.201	&	1.424	&	1.163	&	-0.261	\\
	&			&			&	USDB/CD	&	0.443	&	0.162	&	1.251	&	0.854	&	-0.398	\\
	&			&			&	USD/CD	&	0.423	&	0.142	&	1.973	&	0.937	&	-1.036	\\
	&			&			&	Average :	&	0.424$\pm$0.018	&	0.060$\pm$0.195	&	1.549$\pm$0.377	&	0.984$\pm$0.160	&	-0.565$\pm$0.414	\\
\hline					
$^{26}$Na($\beta^-$)$^{26}$Mg; $^{26}$P($\beta^+$)$^{26}$Si	&	$3^+	2$	&	$2_1^+	1$	&	USDA/CD	&	0.802 	&	0.068	&	0.259	&	3.194	&	2.935	\\
	&		&	&	USDB/CD	&	0.853	&	0.101	&	0.534	&	2.974	&	2.440	\\
	&		&	&	USD/CD	&	0.923	&	0.068	&	0.332	&	2.934	&	2.603	\\
 	&		&	&	Average:&	0.859$\pm$0.061 &	-0.079$\pm$0.019	&	0.375$\pm$0.143	&	3.034$\pm$0.140	&	2.659$\pm$0.252	\\
\hline
$^{28}$Al($\beta^-$)$^{28}$Si; $^{28}$P($\beta^+$)$^{28}$Si	&	$3^+	1$	&	$2_1^+	0$	&	USDA/CD	&	0.695	&	0.126	&	0.406	&	0.703	&	0.297	\\
	&			&			&	USDB/CD	&	0.750	&	0.052	&	0.547	&	0.613	&	0.067	\\
	&			&			&	USD/CD	&	0.810	&	0.053	&	0.566	&	1.188	&	0.622	\\
	&			&			&	Average :	&	0.751$\pm$0.058	&	-0.042$\pm$0.090	&	0.506$\pm$0.087	&	0.835$\pm$0.309	&	0.329$\pm$0.279	\\
\hline				
$^{31}$Al($\beta^-$)$^{31}$Si; $^{31}$Ar($\beta^+$)$^{31}$Cl	&	$\frac{5}{2}^+	\frac{5}{2}$	&	$\frac{3}{2}_1^+	\frac{3}{2}$	&	USDA/CD	&	0.862	&	1.148	&	-1.719	&	7.286	&	9.005	\\
	&			&			&	USDB/CD	&	0.895	&	1.180	&	-1.763	&	7.189	&	8.952	\\
	&			&			&	USD/CD	&	0.894	&	1.246	&	-1.660	&	6.657	&	8.317	\\
	&			&			&	Average :	&	0.884$\pm$0.019	&	-0.426$\pm$1.364	&	-1.714$\pm$0.052	&	7.044$\pm$0.339	&	8.758$\pm$0.383	\\
\hline					
$^{35}$S($\beta^-$)$^{35}$Cl; $^{35}$K($\beta^+$)$^{35}$Ar	&	$\frac{3}{2}^+	\frac{3}{2}$	&	$\frac{3}{2}_1^+	\frac{1}{2}$	&	USDA/CD	&	0.520	&	0.045	&	-0.062	&	1.928	&	1.990	\\
	&			&			&	USDB/CD	&	0.534	&	0.060	&	0.052	&	2.045	&	1.993	\\
	&			&			&	USD/CD	&	0.471	&	0.125	&	0.710	&	2.414	&	1.704	\\
	&			&			&	Average :	&	0.508$\pm$0.033	&	-0.037$\pm$0.093	&	0.233$\pm$0.416	&	2.129$\pm$0.254	&	1.896$\pm$0.166	\\
\hline		
$^{35}$P($\beta^-$)$^{35}$S; $^{35}$Ca($\beta^+$)$^{35}$K	&	$\frac{1}{2}^+	\frac{5}{2}$	&	$\frac{1}{2}_1^+	\frac{3}{2}$	&	USDA/CD	&	0.927	&	0.761	&	-3.339	&	2.994	&	6.333	\\
	&			&			&	USDB/CD	&	0.983	&	0.831	&	-3.322	&	2.755	&	6.077	\\
	&			&			&	USD/CD	&	1.059	&	0.661	&	-2.902	&	2.441	&	5.344	\\
	&			&			&	Average :	&	0.990$\pm$0.067	&	0.244$\pm$0.874	&	-3.188$\pm$0.247	&	2.730$\pm$0.277	&	5.918$\pm$0.513	\\
\end{tabular}
\end{ruledtabular}
\end{table*}

\begin{table*}[ht!]
\caption{Numerical results for the radial mismatch correction are presented. The + (-) sign corresponds to the $\beta^+$ ($\beta^-$) decay of a given pair of Gamow-Teller transitions. Only the effective interactions listed in Table~\ref{tb:2lev} are used in these calculations. The superscript $s$ indicates the inclusion of the surface-peaked term with the Woods-Saxon potential. The correction values are given in percentage (\%).}
\begin{ruledtabular}
\label{resC2}
\begin{tabular}{c|c|c|c|c|c|c|c|c}				
Mirror Pair	&	$J_i^\pi,T_i$	&	$J_f^\pi,T_f$	&	$\delta_{C2}^-$	&	$\delta_{C2}^+$	&	$\delta_{C2}^{s-}$	&	$\delta_{C2}^{s+}$	&	average$^-$	&	average$^+$	\\
\hline				
$^8$Li$(\beta^-)^8$Be; $^8$B($\beta^+$)$^8$Be	&	$2^+,1$	&	$2_1^+,0$	&	6.925	&	11.336	&	9.73	&	13.237	&	1.403$\pm$1.983	&	12.287$\pm$1.344	\\
$^9$Li($\beta^-$)$^9$Be; $^9$C($\beta^+$)$^9$B	&	$\frac{3}{2}^-,\frac{3}{2}$	&	$\frac{3}{2}_1^-,\frac{3}{2}$	&	1.559	&	2.862	&	3.509	&	5.057	&	0.975$\pm$1.379	&	3.960$\pm$1.552	\\
$^{12}$B($\beta^-$)$^{12}$C; $^{12}$N($\beta^+$)$^{12}$C	&	$1^+,1$	&	$0_1^+,0$	&	8.597	&	16.356	&	8.628	&	15.497	&	0.016$\pm$0.022	&	15.927$\pm$0.607	\\
$^{13}$B($\beta^-$)$^{13}$C; $^{13}$O($\beta^+$)$^{13}$N	&	$\frac{3}{2}^-,\frac{3}{2}$	&	$\frac{1}{2}_1^-,\frac{1}{2}$	&	2.464	&	5.777	&	3.75	&	5.713	&	0.643$\pm$0.909	&	5.745$\pm$0.045	\\
$^{17}$N($\beta^-$)$^{17}$O; $^{17}$Ne($\beta^+$)$^{17}$F	&	$\frac{1}{2}^-,\frac{3}{2}$	&	$\frac{3}{2}_1^-,\frac{1}{2}$	&	1.481	&	5.039	&	0.587	&	1.367	&	0.447$\pm$0.632	&	3.203$\pm$2.596	\\
$^{20}$F($\beta^-$)$^{20}$Ne; $^{20}$Na($\beta^+$)$^{20}$Ne	&	$2^+,1$	&	$2_1^+,0$	&	1.106	&	2.787	&	0.808	&	1.888	&	0.149$\pm$0.211	&	2.338$\pm$0.636	\\
$^{20}$O($\beta^-$)$^{20}$F; $^{20}$Mg($\beta^+$)$^{20}$Na	&	$0^+,2$	&	$1_1^+,1$	&	0.368	&	1.626	&	0.56	&	1.517	&	0.096$\pm$0.136	&	1.572$\pm$0.077	\\
$^{21}$F($\beta^-$)$^{21}$Ne; $^{21}$Mg($\beta^+$)$^{21}$Na	&	$\frac{5}{2}^+,\frac{3}{2}$	&	$\frac{3}{2}_1^+,\frac{1}{2}$	&	1.072	&	2.898	&	0.9	&	2.035	&	0.086$\pm$0.122	&	2.467$\pm$0.610	\\
$^{22}$O($\beta^-$)$^{22}$F; $^{22}$Si($\beta^+$)$^{22}$Al	&	$0^+,3$	&	$1_1^+,2$	&	0.468	&	0.276	&	0.068	&	0.467	&	0.268$\pm$0.283	&	0.372$\pm$0.135	\\
$^{24}$Ne($\beta^-$)$^{24}$Na; $^{24}$Si($\beta^+$)$^{24}$Al	&	$0^+,2$	&	$1_1^+,1$	&	0.252	&	1.529	&	0.857	&	2.232	&	0.303$\pm$0.428	&	1.881$\pm$0.497	\\
$^{25}$Na($\beta^-$)$^{25}$Mg; $^{25}$Si($\beta^+$)$^{25}$Al	&	$\frac{5}{2}^+,\frac{3}{2}$	&	$\frac{5}{2}_1^+,\frac{1}{2}$	&	0.545	&	0.301	&	1.408	&	3.879	&	0.432$\pm$0.610	&	2.090$\pm$2.530\\
$^{26}$Na($\beta^-$)$^{26}$Mg; $^{26}$P($\beta^+$)$^{26}$Si	&	$3^+,2$	&	$2_1^+,1$	&	0.3994 &	11.961	&	3.897	&	9.199	&	3.946$\pm$0.069	&	10.580$\pm$1.953 \\ 
$^{28}$Al($\beta^-$)$^{28}$Si; $^{28}$P($\beta^+$)$^{28}$Si	&	$3^+,1$	&	$2_1^+,0$	&	2.558	&	10.285	&	1.297	&	6.312	&	0.631$\pm$0.892	&	8.299$\pm$2.809	\\
$^{31}$Al($\beta^-$)$^{31}$Si; $^{31}$Ar($\beta^+$)$^{31}$Cl	&	$\frac{5}{2}^+,\frac{5}{2}$	&	$\frac{3}{2}_1^+,\frac{3}{2}$	&	1.261	&	4.527	&	1.609	&	0.456	&	0.174$\pm$0.246	&	2.492$\pm$2.879\\
$^{35}$S($\beta^-$)$^{35}$Cl; $^{35}$K($\beta^+$)$^{35}$Ar	&	$\frac{3}{2}^+,\frac{3}{2}$	&	$\frac{3}{2}_1^+,\frac{1}{2}$	&	0.187	&	3.607	&	0.311	&	1.692	&	0.062$\pm$0.088	&	2.650$\pm$1.354	\\
$^{35}$P($\beta^-$)$^{35}$S; $^{35}$Ca($\beta^+$)$^{35}$K	&	$\frac{1}{2}^+,\frac{5}{2}$	&	$\frac{1}{2}_1^+,\frac{3}{2}$	&	0.531	&	5.875	&	1.185	&	4.175	&	0.327$\pm$0.462	&	5.025$\pm$1.202	\\
\end{tabular}
\end{ruledtabular}
\end{table*}

\begin{table*}
\caption{Numerical results for the higher-order isospin-symmetry-breaking correction terms are presented. The + (-) sign corresponds to the $\beta^+$ ($\beta^-$) decay of a given pair of Gamow-Teller transitions. Only the effective interactions listed in Table~\ref{tb:2lev} are used in these calculations. The surface-peaked term of the Woods-Saxon potential is not considered for $\delta_{C3}^A$ calculations. The correction values are given in percentage (\%).}
\begin{ruledtabular}
\begin{tabular}{c|c|c|c|c|c|c|c|c|c|c}
Mirror Pair	&	$J_i^\pi,T_i$	&	$J_f^\pi,T_f$	&	$\delta_{C3}^-$	&	$\delta_{C3}^+$	&	$\delta_{C4}^-$	&	$\delta_{C4}^+$	&	$\delta_{C5}^-$	&	$\delta_{C5}^+$	&	$\delta_{C6}^-$	&	$\delta_{C6}^+$	\\
\hline			
$^8$Li$(\beta^-)^8$Be; $^8$B($\beta^+$)$^8$Be	&	$2^+,1$	&	$2_1^+,0$	&	0.673	&	2.751	&	-0.037	&	-0.035	&	0.013	&	-0.051	&	-0.001	&	-0.019	\\
$^9$Li($\beta^-$)$^9$Be; $^9$C($\beta^+$)$^9$B	&	$\frac{3}{2}^-,\frac{3}{2}$	&	$\frac{3}{2}_1^-,\frac{3}{2}$	&	0.227	&	0.233	&	-0.026	&	-0.026	&	0.004	&	0.004	&	0	&	0	\\
$^{12}$B($\beta^-$)$^{12}$C; $^{12}$N($\beta^+$)$^{12}$C	&	$1^+,1$	&	$0_1^+,0$	&	0.279	&	0.288	&	-0.022	&	-0.327	&	-0.004	&	-0.016	&	0	&	0	\\
$^{13}$B($\beta^-$)$^{13}$C; $^{13}$O($\beta^+$)$^{13}$N	&	$\frac{3}{2}^-,\frac{3}{2}$	&	$\frac{1}{2}_1^-,\frac{1}{2}$	&	0.096	&	0.104	&	-0.003	&	-0.046	&	0.001	&	-0.002	&	0	&	0	\\
$^{17}$N($\beta^-$)$^{17}$O; $^{17}$Ne($\beta^+$)$^{17}$F	&	$\frac{1}{2}^-,\frac{3}{2}$	&	$\frac{3}{2}_1^-,\frac{1}{2}$	&	-0.166	&	-0.718	&	-0.578	&	-7.446	&	0.013	&	0.196	&	0	&	-0.001	\\
$^{20}$F($\beta^-$)$^{20}$Ne; $^{20}$Na($\beta^+$)$^{20}$Ne	&	$2^+,1$	&	$2_1^+,0$	&	-0.004	&	-0.074	&	-0.132	&	-0.585	&	0	&	0.006	&	0	&	0	\\
$^{20}$O($\beta^-$)$^{20}$F; $^{20}$Mg($\beta^+$)$^{20}$Na	&	$0^+,2$	&	$1_1^+,1$	&	-0.013	&	-0.076	&	-0.223	&	-0.419	&	0.001	&	0.005	&	0	&	0	\\
$^{21}$F($\beta^-$)$^{21}$Ne; $^{21}$Mg($\beta^+$)$^{21}$Na	&	$\frac{5}{2}^+,\frac{3}{2}$	&	$\frac{3}{2}_1^+,\frac{1}{2}$	&	-0.052	&	-0.267	&	-0.124	&	-0.341	&	0.002	&	0.016	&	0	&	0	\\
$^{22}$O($\beta^-$)$^{22}$F; $^{22}$Si($\beta^+$)$^{22}$Al	&	$0^+,3$	&	$1_1^+,2$	&	0.471	&	4.368	&	-67.581	&	-41.057	&	-0.387	&	-2.799	&	-0.001	&	-0.048	\\
$^{24}$Ne($\beta^-$)$^{24}$Na; $^{24}$Si($\beta^+$)$^{24}$Al	&	$0^+,2$	&	$1_1^+,1$	&	-0.016	&	-0.056	&	-0.023	&	-0.098	&	0	&	0.002	&	0	&	0	\\
$^{25}$Na($\beta^-$)$^{25}$Mg; $^{25}$Si($\beta^+$)$^{25}$Al	&	$\frac{5}{2}^+,\frac{3}{2}$	&	$\frac{5}{2}_1^+,\frac{1}{2}$	&	-0.014	&	0.001	&	-0.009	&	-0.003	&	0	&	0	&	0	&	0	\\
$^{26}$Na($\beta^-$)$^{26}$Mg; $^{26}$P($\beta^+$)$^{26}$Si	&	$3^+,2$	&	$2_1^+,1$	&	0.004	&	-0.021	&	-0.047	&	-0.555	&	0	&	0.002	&	0	&	0	\\
$^{28}$Al($\beta^-$)$^{28}$Si; $^{28}$P($\beta^+$)$^{28}$Si	&	$3^+,1$	&	$2_1^+,0$	&	-0.004	&	-0.013	&	-0.024	&	-0.297	&	0	&	0.001	&	0	&	0	\\
$^{31}$Al($\beta^-$)$^{31}$Si; $^{31}$Ar($\beta^+$)$^{31}$Cl	&	$\frac{5}{2}^+,\frac{5}{2}$	&	$\frac{3}{2}_1^+,\frac{3}{2}$	&	0.023	&	-0.111	&	-0.001	&	-0.343	&	0	&	0.007	&	0	&	0	\\
$^{35}$S($\beta^-$)$^{35}$Cl; $^{35}$K($\beta^+$)$^{35}$Ar	&	$\frac{3}{2}^+,\frac{3}{2}$	&	$\frac{3}{2}_1^+,\frac{1}{2}$	&	-0.018	&	-0.431	&	0	&	-0.08	&	0	&	0.012	&	0	&	0	\\
$^{35}$P($\beta^-$)$^{35}$S; $^{35}$Ca($\beta^+$)$^{35}$K	&	$\frac{1}{2}^+,\frac{5}{2}$	&	$\frac{1}{2}_1^+,\frac{3}{2}$	&	0.017	&	-0.081	&	-0.019	&	-0.186	&	0	&	0.003	&	0	&	0	\\
\end{tabular}
\end{ruledtabular}
\label{higher}
\end{table*}

\begin{table*}
\caption{Numerical results for the mirror asymmetry parameter of the GT transitions are shown. The individual contributions of each isospin-symmetry-breaking (ISB) correction term are also provided, with $\delta_5^{ISB}$ and $\delta_6^{ISB}$ being negligible. $\delta^{ISB}$ is the total ISB contribution. The values are given in percentage (\%). The table includes the results of the previous shell model calculations SmVo2003~\cite{Smirnova2003}. The experimental data for $A=24$ are taken from Ref.~\cite{Smirnova2003,PhysRevC.80.044302,sym13122278}.}
\label{delta}
\begin{ruledtabular}
\begin{tabular}{c|c|c|c|c|c|c|c|c|c}
Mirror Pair	&	$J_i^\pi,T_i$	&	$J_f^\pi,T_f$	&	$\delta_1^{ISB}$	&	$\delta_2^{ISB}$	&	$\delta_3^{ISB}$	&	$\delta_4^{ISB}$	&	$\delta^{ISB}$	&	SmVo2003 &	$\delta^{exp}$	\\
\hline															
$^8$Li$(\beta^-)^8$Be;	$^8$B($\beta^+$)$^8$Be	&	$2^+,1$	&	$2_1^+,0$	&	2.482$\pm$0.433	&	10.884$\pm2.396$	&	2.078	&	0.002	&	15.364$\pm$2.435	&	18.82	&	8.4$\pm$1.8 	\\
$^9$Li($\beta^-$)$^9$Be;	$^9$C($\beta^+$)$^9$B	&	$\frac{3}{2}^-,\frac{3}{2}$	&	$\frac{3}{2}_1^-,\frac{3}{2}$	&	-0.582$\pm$1.036	&	2.985$\pm$2.076	&	0.006	&	0	&	2.408$\pm$2.320	&	2.72	&	-0.854$\pm$3.382 	\\
$^{12}$B($\beta^-$)$^{12}$C;	$^{12}$N($\beta^+$)$^{12}$C	&	$1^+,1$	&	$0_1^+,0$	&	0.903$\pm$0.215	&	15.911$\pm$0.608	&	0.009	&	-0.305	&	16.506$\pm$0.645	&	21.85	&	12.257$\pm$4.878 	\\
$^{13}$B($\beta^-$)$^{13}$C;	$^{13}$O($\beta^+$)$^{13}$N	&	$\frac{3}{2}^-,\frac{3}{2}$	&	$\frac{1}{2}_1^-,\frac{1}{2}$	&	2.401$\pm$0.293	&	5.102$\pm$0.910	&	0.008	&	-0.043	&	7.465$\pm$0.956	&	8.65	&	12.202$\pm$1.308 	\\
$^{17}$N($\beta^-$)$^{17}$O;	$^{17}$Ne($\beta^+$)$^{17}$F	&	$\frac{1}{2}^-,\frac{3}{2}$	&	$\frac{3}{2}_1^-,\frac{1}{2}$	&	35.834	&	2.756$\pm$2.672	&	-0.552	&	-6.868	&	31.352$\pm$2.672	&	64.8	&	44.544$\pm$2.335 	\\
$^{20}$F($\beta^-$)$^{20}$Ne;	$^{20}$Na($\beta^+$)$^{20}$Ne	&	$2^+,1$	&	$2_1^+,0$	&	6.36	&	2.188$\pm$0.670	&	-0.07	&	-0.453	&	8.032$\pm$0.670	&	1.11	&	1.8$\pm$0.7 	\\
$^{20}$O($\beta^-$)$^{20}$F;	$^{20}$Mg($\beta^+$)$^{20}$Na	&	$0^+,2$	&	$1_1^+,1$	&	2.241	&	1.475$\pm$0.156	&	-0.063	&	-0.196	&	3.462$\pm$0.156	&	-1.53	&	9.875$\pm$1.659 	\\
$^{21}$F($\beta^-$)$^{21}$Ne;	$^{21}$Mg($\beta^+$)$^{21}$Na	&	$\frac{5}{2}^+,\frac{3}{2}$	&	$\frac{3}{2}_1^+,\frac{1}{2}$	&	2.808	&	2.381$\pm$0.622	&	-0.215	&	-0.217	&	4.771$\pm$0.622	&	-0.56	&		-58.313$\pm$7.422\\

$^{22}$O($\beta^-$)$^{22}$F;	$^{22}$Si($\beta^+$)$^{22}$Al	&	$0^+,3$	&	$1_1^+,2$	&	-36.367	&	6.521$\pm$0.044	&	3.897	&	26.523	&	-8.301$\pm$0.044	&		&		209$\pm$96 \\

$^{24}$Ne($\beta^-$)$^{24}$Na;	$^{24}$Si($\beta^+$)$^{24}$Al	&	$0^+,2$	&	$1_1^+,1$	&	18.802	&	1.578$\pm$0.656	&	-0.04	&	-0.075	&	20.267$\pm$0.656	&		&	28.2$\pm$1.8	\\
$^{25}$Na($\beta^-$)$^{25}$Mg;	$^{25}$Si($\beta^+$)$^{25}$Al	&	$\frac{5}{2}^+,\frac{3}{2}$	&	$\frac{5}{2}_1^+,\frac{1}{2}$	&	-0.565$\pm$0.414	&	1.658$\pm$2.603	&	0.015	&	0.006	&	1.115$\pm$2.635	&	1.11	&	-1.599$\pm$11.900	\\
$^{26}$Na($\beta^-$)$^{26}$Mg;	$^{26}$P($\beta^+$)$^{26}$Si	&	$3^+,2$	&	$2_1^+,1$	&	2.659$\pm$0.252	&	6.634$\pm$1.954	&	-0.025	&	-0.508	&	8.760$\pm$1.970	&	&	34.834$\pm$16.181\\
$^{28}$Al($\beta^-$)$^{28}$Si;	$^{28}$P($\beta^+$)$^{28}$Si	&	$3^+,1$	&	$2_1^+,0$	&	0.329$\pm$0.279	&	7.668$\pm$2.947	&	-0.009	&	-0.273	&	7.716$\pm$2.961	&	11.54	&	-3.706$\pm$0.444	\\
$^{31}$Al($\beta^-$)$^{31}$Si;	$^{31}$Ar($\beta^+$)$^{31}$Cl	&	$\frac{5}{2}^+,\frac{5}{2}$	&	$\frac{3}{2}_1^+,\frac{3}{2}$	&	8.758$\pm$0.383	&	2.317$\pm$2.889	&	-0.134	&	-0.342	&	10.607$\pm$2.914	&	16.27	&	33$\pm$18	\\
$^{35}$S($\beta^-$)$^{35}$Cl;	$^{35}$K($\beta^+$)$^{35}$Ar	&	$\frac{3}{2}^+,\frac{3}{2}$	&	$\frac{3}{2}_1^+,\frac{1}{2}$	&	1.896$\pm$0.166	&	2.587$\pm$1.357	&	-0.413	&	-0.08	&	4.003$\pm$1.367	&	4.96	&	14.533$\pm$10.308	\\
$^{35}$P($\beta^-$)$^{35}$S;	$^{35}$Ca($\beta^+$)$^{35}$K	&	$\frac{1}{2}^+,\frac{5}{2}$	&	$\frac{1}{2}_1^+,\frac{3}{2}$	&	5.918$\pm$0.513	&	4.698$\pm$1.288	&	-0.098	&	-0.167	&	10.354$\pm$1.386	&	18.81	&	18$\pm$5 	\\
\end{tabular}
\end{ruledtabular}
\end{table*}

\end{document}